
\input phyzzx
\font\el=cmbx10 scaled \magstep2

\def\L{{\cal L}}
\def\tr{{\rm tr}}
\def\ri{\rightarrow}

\def\Q{{\cal Q}}
\def\A{{\cal A}}
\def\ol{\overleftarrow}

\def\l{\left}
\def\r{\right}

{\obeylines
\hfill CLNS 93/1192
\hfill IP-ASTP-02-93
\hfill ITP-SB-93-04
\hfill August, 1993}

\centerline {{\el Corrections to Chiral Dynamics of Heavy Hadrons:}}
\centerline {{\el (I) 1/M Correction}}

\medskip

\centerline{\bf Hai-Yang Cheng$^{a,d}$, Chi-Yee Cheung$^a$, Guey-Lin Lin$^a$,}
\centerline{\bf Y. C. Lin$^b$, Tung-Mow Yan$^{c}$, and Hoi-Lai Yu$^a$}

\medskip
\centerline{$^a$ Institute of Physics, Academia Sinica, Taipei,}
\centerline{Taiwan 11529, Republic of China}

\medskip
\centerline{$^b$ Physics Department, National Central University,
Chung-li,}
\centerline{Taiwan 32054, Republic of China}

\medskip
\centerline{$^c$ Floyd R. Newman Laboratory of Nuclear Studies, Cornell
University}
\centerline{Ithaca, New York 14853, USA }

\medskip
\centerline{$^d$ Institute for Theoretical Physics, State University of
New York}
\centerline{Stony Brook, New York 11794, USA}

\medskip

\centerline{\bf Abstract}

In earlier publications we have analyzed the strong and radiative
decays of heavy hadrons in a formalism which
incorporates both heavy-quark and chiral symmetries. In particular, we have
derived a heavy-hadron chiral Lagrangian whose coupling constants are related
by the heavy-quark flavor-spin symmetry arising from the QCD Lagrangian with
infinitely massive quarks. In this paper, we re-examine the structure of the
above chiral Lagrangian by including the effects of $1/m_Q$ corrections in
the heavy quark effective theory. The relations among the coupling
constants, originally derived in the heavy-quark limit, are modified
by heavy quark symmetry breaking interactions in QCD. Some of the
implications are discussed.

\endpage

\noindent{\bf 1.  Introduction  }

  In this and a subsequent paper, we would like to examine various symmetry
breaking corrections to the strong and electromagnectic decays of heavy
hadrons. There are two different kinds of symmetry breaking effects on the
chiral dynamics of heavy hadrons: the $1/m_Q$ corrections from the heavy
quarks and the finite-mass effects from the light quarks. We will
focus on the $1/m_Q$ corrections in this work and leave the discussion on
SU(3) breaking effects to the forthcoming paper [1].

As is well known, the QCD dynamics in the limit of infinite quark mass
exhibits a new spin-flavor symmetry which is known as the heavy quark
symmetry (HQS) [2,3]. Corrections
to this symmetry limit can be systematically incorporated
 into the heavy quark effective theory (HQET) of QCD where symmetry
breaking effects are summarized by higher-dimensional operators suppressed
by powers of $1/m_Q$ [4-9].  Such an effective theory has been
a powerful tool to analyze weak-transition form factors of heavy
hadrons containing one single heavy quark [10]. We have recently,
among others,
initiated a study of strong and radiative decays of heavy hadrons by
deriving a heavy-hadron chiral Lagrangian which obeys constraints from the
 heavy quark symmetry [11-17].
As the idea of synthesizing the heavy-quark and chiral symmetries receives
growing attention, there remain important issues to be explored. Especially,
implications of the aforementioned $1/m_Q$
corrections to the structure of the heavy-hadron chiral Lagrangian
have not been systematically studied [18]. Since the charmed quark is
not particularly heavy compared to the QCD scale, such corrections can be
important in the chiral Lagrangian for charmed hadrons.

   As an example to illustrate the issues involved, consider the heavy-meson
chiral Lagrangian given by Eq.(2.16) of Ref.[17]:
$$\eqalign{{\cal L}^{(1)}_{PP^*}&=D_{\mu}PD^{\mu}P^{\dagger}-M_P^2
PP^{\dagger}+f\sqrt{M_PM_{P^*}}\,(P\A^{\mu}P_{\mu}^{*\dagger}
+P_{\mu}^* \A^{\mu}P^{\dagger})\cr
&-{1\over 2}P^{*\mu\nu}P_{\mu\nu}^{*\dagger}+M_{P^*}^2P^{*\mu}
P_{\mu}^{*\dagger}\cr
&+{1\over 2}g\epsilon_{\mu\nu\lambda\kappa}
(P^{*\mu\nu} \A^{\lambda}P^{*\kappa\dagger}+P^{*\kappa}
\A^{\lambda}P^{*\mu\nu\dagger}),\cr}\eqno(1.1)$$
where $P$ and $P^*$ are the ground-state heavy mesons with quantum numbers
$J^P=0^-$ and $1^-$ respectively, and
$$P^{*\dagger}_{\mu\nu}=D_{\mu}P^{*\dagger}_{\nu}-
D_{\nu}P^{*\dagger}_{\mu},
\eqno(1.2a)$$
$$D_{\mu}P^{*\dagger}_\nu=\partial_{\mu}P^{*\dagger}_\nu+{\cal V}_{\mu}P^{
\dagger}_\nu-ieA_{\mu}(P^{*\dagger}_\nu{\cal Q'}-{\cal Q}P^{*\dagger}_\nu),
\eqno(1.2b)$$
and a similar definition for the covariant derivative $D_{\mu}P^{\dagger}$.
In Eq.(1.1)
$A_{\mu}$ is the electromagnetic field whereas ${\cal V}_{\mu}$ and ${\cal A}_
{\mu}$ are respectively the chiral vector and chiral axial fields (see
Ref.[17] for more detail). The prediction from heavy quark symmetry consists
of two parts. The flavor symmetry implies that the coupling constants $f$ and
$g$ are the same for any heavy flavor. The spin symmetry relates the two
parameters by
$$g={1\over 2}f.\eqno(1.3)$$
Similar predictions have also been obtained for the heavy baryon chiral
Lagrangian.
These predictions help reduce the number of unknowns in the heavy-hadron
chiral Lagrangian. For instance, the $D^*D\pi$ and $D^*D\gamma$ coupling
constants are related to those of $D^*D^*\pi$ and $D^*D^*\gamma$
respectively. This is crucial
since the latter two couplings are very difficult to measure in practice. With
the knowledge of above coupling strengths, the predictive power of the heavy
meson chiral Lagrangian is greatly enhanced. However, success of such a scheme
demands an assessment of how large the $1/m_Q$ corrections are. The purpose
of this paper is to study such type of $1/m_Q$ corrections which modify the
various HQS relations among the coupling constants. At present, we do not
attempt to give a quantitative predictions on the
sizes of various $1/m_Q$ effects.
A quantitative analysis will be presented in a future publication.

   As is well-known, there are two energy scales in the chiral perturbation
theory involving a heavy hadron: the mass of the heavy hadron $M_H$ and the
the chiral symmetry breaking
scale $\Lambda_{\chi}$. In principle, one may expand the theory in inverse
powers of these two scales. However, because the heavy hadrons have large
masses, the derivatives acting on the heavy hadron fields will produce
large momentum factors. This complicates the power counting procedure. This
difficulty is overcome by a simple observation. Strong and electromagnetic
interactions at low energies of a heavy hadron with other light hadrons are
governed by the
energy scale $\Lambda_{\rm QCD}$ which is much smaller than $M_H$.
Consequently, the four momentum of a heavy hadron has only fluctuations of
the order of $\Lambda_{\rm QCD}$ throughtout its history. Its momentum can,
therefore, be parametrized as
$$P=\,M_Hv+k,~~~v^2=1,\eqno(1.4)$$
where $k$ is of order $\Lambda_{\rm QCD}$. In accordance with the
parametrization, one introduces a velocity-dependent field $H_v(x)$ by [12,19]
$$H(x)=e^{-iM_Hv\cdot x}H_v(x),\eqno(1.5)$$
where $H(x)$ is the standard field operator for a heavy hadron. The
velocity-dependent field $H_v(x)$ carries only the residual momentum $k$. It
follows from (1.5) that
$$\partial_{\mu}H(x)=e^{-iM_Hv\cdot x}[-iM_Hv_{\mu}H_v(x)+
\partial_{\mu}H_v(x)].\eqno(1.6)$$
The dependence on the large mass $M_H$ is now made explicit: the second term
in (1.6) is of order $k/M_H$ relative to the first one. In terms of $H_v(x)$,
derivatives acting on the heavy hadron and Goldstone boson fields are treated
on equal footing, and a consistent $1/M_H$ and $1/\Lambda_\chi$ expansion can
be developed for the heavy hadron chiral Lagrangian.

The velocity-dependent fields for the $0^-$ and $1^-$ heavy hadrons of (1.1)
are
$$P(x)=e^{-iM_{P^*}v\cdot x}P_v(x),\eqno(1.7a)$$
$$P^*_{\mu}(x)=e^{-iM_{P^*}v\cdot x}P^*_{v,\mu}(x),\eqno(1.7b)$$
$$v\cdot P^*(v)=0.\eqno(1.7c)$$
To simplify our notation, we write $P(v)\equiv P_v(x)$ and
$P^*_{\mu}(v)\equiv P^*_{v,\mu}(x)$.
Retaining only the leading terms, we obtain
$$\eqalign{{\cal L}_{v,PP^*}^{(1)}
&=-2iM_{P^*}P(v)v\cdot DP^{\dagger}(v)+2iM_{P^*}
P^{*\mu}(v)v\cdot D P_{\mu}^{*\dagger}(v)\cr
&+\Delta M^2P(v)P^{\dagger}(v) +f\sqrt{M_PM_{P^*}}\,[P(v)\A^{\mu}P_{\mu}^{*
\dagger}(v)+P_{\mu}^*(v) \A^{\mu}P^{\dagger}(v)]\cr
&+2iM_{P^*}g\epsilon_{\mu\nu\lambda\kappa}
P^{*\mu}(v)v^{\nu}\A^{\lambda}P^{*\kappa\dagger}(v), \cr}\eqno(1.8)$$
with
$$\Delta M^2=M_{P^*}^2-M_P^2.\eqno(1.9)$$
Note that we have neglected terms which are suppressed by  $1/M_{
P^*}$ comparing with the leading contributions. Therefore, ${\cal L}_{v,PP^*}
^{(1)}$
is the leading-order heavy-meson chiral Lagrangian in the double expansions
of $1/M_{P^*}$ and Goldstone-boson momenta.
Before proceeding further, we should like to make two remarks on the
Lagrangian ${\cal L}_{v,PP^*}^{(1)}$. First of all, the parameters $M_P$ and
$M_{P^*}$ in Eq. (1.8) are taken to be the physical masses of the heavy
mesons $P$ and $P^*$ respectively. This accounts for the appearence of $\Delta
 M^2P(v)P^{\dagger}(v)$ in Eq. (1.8). Theoretically, we expect
$$M_{P^*}-M_P=\,{\cal O}\left({\Lambda_{\rm QCD}^2\over m_Q}\right),
\eqno(1.10)$$
so $\Delta M^2$ is of order $\Lambda_{\rm QCD}^2$
and it is a simplest $1/M_H$
correction to the leading terms of ${\cal L}_{v,PP^*}^{(1)}$ which we keep.
Second, the coupling constants $f$ and $g$ are no longer assumed to satisfy
the spin symmetry relation (1.3).

   It has been noted by Luke and Manohar [20] that the structure of the
$1/M_H$ expansion must satisfy the ``reparametrization invariance'' which
is a conseqence of the nonuniqueness of the parametrization (1.4). The
four-velocity $v$ and the residual momentum $k$ can be arbitrarily chosen so
long as $v^2=1$ and $k\sim\Lambda_{\rm QCD}<<M_H$.
For consistency, the heavy-meson chiral theory must be invariant under the
transformation
$$v\rightarrow v+r/M _{P^*}, \ \ \ k\rightarrow k-r,\eqno(1.11a)$$
$$(v+r/M_{P^*})^2=1.\eqno(1.11b)$$
This leads to the conclusion that a reparametrization-invariant heavy-meson
chiral Lagrangian, which we denote as ${\tilde {\cal L}}_{
PP^*}$, must have the following structure [20]
$${\tilde {\cal L}}_{PP^*}=\sum_{v}{\tilde {\cal L}}_{v,PP^*}(P(v), {\tilde P}
^{*\mu}(v), {\cal W}^{\mu}),\eqno(1.12)$$
where
$${\cal W}_{\mu}=v_{\mu}+{iD_{\mu}\over M_{P^*}},\eqno(1.13a)$$
$${\tilde P}^{*\mu}(v)=P^{*\mu}(v)-v^{\mu}{iD\cdot P^*(v)\over M_{P^*}}.
\eqno(1.13b)$$
Therefore, to maintain the reparametrization invariance to order
$O(1/M_{P^*})$, the Lagrangian is agumented to be
$$\eqalign{{\tilde {\cal L}}_{v,PP^*}^{(1)}=&-M_{P^*}^2P(v)({\cal W}^2-1)P
^{\dagger}(v)+M_{P^*}^2{\tilde P}^{*\mu}(v)({\cal W}^2-1) {\tilde P}_{\mu}^{*
\dagger}(v) \cr
&+\Delta M^2P(v)P^{\dagger}(v)
+f\sqrt{M_PM_{P^*}}\,[P(v)\A^{\mu}{\tilde P}_{\mu}^{*\dagger}(v)+{\tilde P}_{
\mu}^*(v) \A^{\mu}P^{\dagger}(v)] \cr
&+iM_{_{P^*}}g\epsilon_{\mu\nu\lambda\kappa}[{\tilde P}^{*\mu}(v)
{\overleftarrow{{\cal W}}}{^{\nu}}\A^{\lambda}{\tilde P}^{*
\kappa\dagger}(v)+{\tilde P}^{*\mu}(v)\A^{\lambda}{\cal W}^\nu{\tilde P}^{*
\kappa\dagger}(v)].   \cr}\eqno(1.14)$$
The prescription (1.13) uniquely determines the terms of order $1/M_{P^*}$
necessary to ensure the reparametrization invariance of ${\tilde {\cal L}}_{
v,PP^*}^{(1)}$. These $1/M_{P^*}$ corrections are essentially kinematic in
nature. They are important, but they can be retrieved by following the
prescription (1.13). However, there are other $1/M_{P^*}$ contributions which
are reparametrization invariant by themselves, but at least contain two
derivatives. It should be pointed out that the original Lagrangian
(1.1) is reparametrization invariant. Eq.(1.14) follows simply from Eq.(1.1)
by keeping the first two leading orders in the $1/M_H$
expansion using Eq.(1.7)
(in particular, Eq.(1.7c) should hold to order $1/M_H$). The requirement of
reparametrization invariance will become more useful as the $1/M_H$ expansion
is carried out to higher orders, or when we deal with new situations [21].

  There is another type of $1/M_{P^*}$ corrections which will be the focus
of the present work. In contrast with the previous corrections, these are
 dynamical in nature and they arise from taking into account the $1/m_Q$ terms
 in HQET. It is well known that the following
two operators in HQET break the heavy-quark spin-flavor symmetry at the order
of $1/m_Q$ [7,8]:
$${\cal L}=\bar{h}_viv\cdot Dh_v+{\cal L}_I,\eqno(1.15a)$$
$${\cal L}_I=\,O_1+O_2,\eqno(1.15b)$$
$$O_1={1\over 2m_Q}\bar{h}_v(iD)^2h_v,\eqno(1.15c)$$
$$O_2={1\over 2m_Q}\bar{h}_v(-{1\over 2}g_s\sigma_{\mu\nu}G^{\mu\nu})
h_v.\eqno(1.15d)$$
Specifically, the operator $O_1$ breaks the flavor symmetry, and the operator
$O_2$ breaks both the flavor and spin symmetries. To the first order in the
Goldstone boson's momentum, the only effects of $O_1$ and $O_2$ are to make
$1/m_Q$ corrections to the coupling constants $f$ and $g$ which appear in
Eq.(1.14). To order $1/m_Q$, we may write
$$f=f_0+f_c{\Lambda\over 2m_Q},\eqno(1.16a)$$
$$g=g_0+g_c{\Lambda\over 2m_Q},\eqno(1.16b)$$
where $g_0={1\over 2}f_0$, and $\Lambda$ is an arbitrary mass scale.
Presumably, the value for $\Lambda$ should be chosen in such a way that $f_c
\approx f_0$ and $g_c\approx g_0$. Under this requirement, it has recently
been argued that the parameter $\Lambda$ is of order
$\Lambda_{\chi}$ rather than $\Lambda_{\rm
QCD}$ [22]. However, we will take no position on this point as it is still
not widely accepted.

   We observe that the two types of $1/M_H\,(1/m_Q)$ corrections discussed
above have distinct characteristics. The $1/M_H$ correction demanded by
reparametrization invariance
introduces new structures which modify the leading
order Lagrangian. The other, dynamical corrections of order $1/m_Q$ produce
heavy quark symmetry breaking contributions to the coupling constants in
the leading order Lagrangian but they do not alter the structure of the
Lagrangian. The two effects together provide the complete $1/M_H\,(1/m_Q)$
corrections to heavy quark symmetry. Since $1/M_H=1/m_Q+O(1/m_Q^2)$,
there is no need to keep the difference between $1/M_H$ and $1/m_Q$ at this
order.

   So far, we have used the heavy meson dynamics as an example to discuss the
various issues
in the $1/M_H\,(1/m_Q)$ corrections to the heavy quark symmetry.
Clearly, we can carry out a similar discussion for heavy baryons on
reparametrization invariance and dynamical corrections to the coupling
constants.

   With the issues in the $1/M_H\,(1/m_Q)$ corrections clearly defined,
we will concentrate our attention in what follows on interactions between the
heavy hadrons and the Goldstone bosons with a single derivative.
Section 2 is devoted to a study of the $O(1/m_Q)$ correction to the coupling
constants for both strong and electromagnetic interactions in the heavy
meson secetor. A similar study for heavy baryons is carried out in Section 3.
We will employ the method of interpolating fields extensively utilized in
Ref.[17]. We find that all the heavy quark spin symmetry relations among
the coupling constants (both strong and electromagnetic) are completely broken
by $1/m_Q$ corrections.

Finally, in Section 4 we make some concluding remarks and we shall comment
on the work done by Randall and Sather
[23] concerning the $SU(3)$-violating corrections to the  heavy-meson
hyperfine splitting, which is a typical $O(1/m_Q)$ phenomenon.
As we shall point out, the calculation performed in Ref. [23] is incomplete,
namely it does not include all the corrections of order $1/m_Q$.

\endpage

\noindent{\bf 2.~~~$1/m_Q$ Corrections to the
Dynamics of Heavy Mesons}
\vskip 0.3 cm

In this section we shall study the $1/m_Q $ corrections to the coupling
constants of the heavy-meson
chiral Lagrangian given by Eqs. (2.16) and (2.19) of Ref.[17].
First of all, we shall rewrite the chiral Lagrangians
$\L^{(1)}_{PP^*}$ and $\L^{(2)}_{PP^*}$ in terms of velocity-dependent fields
and retain only the leading terms in the $1/M_H$ expansion.
The velocity-dependent version of $\L^{(1)}_{PP^*}$ is given by Eq. (1.8),
which we recall here for convenience:
$$\eqalign{\L^{(1)}_{v,PP^*}&=-2iM_{P^*}P(v)v\cdot DP^{\dagger}(v)+2iM_{P^*}P^
{*\mu}(v)v\cdot D P_{\mu}^{*\dagger}(v) \cr
&+\Delta M^2P(v)P^{\dagger}(v)
+f\sqrt{M_PM_{P^*}}\,[P(v){\cal A}^{\mu}P_{\mu}^
{*\dagger}(v)+P_{\mu}^*(v) {\cal A}^{\mu}P^{\dagger}(v)] \cr
&+2iM_{P^*}\,g\epsilon_{\mu\nu\lambda\kappa}
P^{*\mu}(v)v^{\nu}{\cal A}^{\lambda}P^{*\kappa\dagger}(v), \cr}\eqno(2.1)$$
with
$$\Delta M^2=M_{P^*}^2-M_P^2.\eqno(2.2)$$
Substituting Eqs.(1.7a) and (1.7b) into $\L^{(2)}_{PP^*}$ which describes
the radiative transitions, we obtain (see
Ref.[17] for notations)
$$\eqalign{\L^{(2)}_{v,PP^*}&=\sqrt{M_PM_{P^*}}\epsilon_{\mu\nu
\alpha\beta}v^{\alpha}P^{*\beta}(v)[{1\over 2}d(\xi^{\dagger}
{\cal Q}\xi
+\xi{\cal Q}\xi^{\dagger})+d^{\prime}{\cal Q}^{\prime}]
F^{\mu\nu}P^{\dagger}(v)+h.c.\cr
&+id^{''}M_{P^*}F_{\mu\nu}P^{*\nu}(v)[\gamma {\cal Q}^{\prime}
-{1\over 2}(\xi^{\dagger}{\cal Q}\xi
+\xi{\cal Q}\xi^{\dagger})]P^{*\mu\dagger}(v),\cr}\eqno(2.3)$$
where ${\cal Q}'$ is the heavy-quark's charge and ${\cal Q}$
denotes the charge matrix of the light quarks:
$${\cal Q}=\pmatrix{{2\over 3}&0&0\cr
0&-{1\over 3}& 0\cr
0&0&-{1\over 3}\cr}.\eqno(2.4)$$
Note that, contrary to Eq. (2.19) of Ref.[17], we do not need to subtract
from ${\cal L}_{v,PP^*}^{(2)}$ the normal magnetic moment term of $P^*_{\mu}$
induced by the minimum substitution. This is because such contributions are
not among the leading terms kept in (2.3).

As indicated in Eqs.(1.16a) and (1.16b), every coupling constant in $\L^{(1)}_
{v,PP^*}$ and $\L^{(2)}_{v,PP^*}$ can be expanded in powers of $1/m_Q$. In
particular, we have written there the expansion for
coupling constants $f$ and $g$:
$$f=f_0+f_c{\Lambda\over 2m_Q},\eqno(2.5)$$
and
$$g=g_0+g_c{\Lambda\over 2m_Q}.\eqno(2.6)$$
The zeroth order contributions $f_0$ and $g_0$ are related by HQS [11], namely
$$g_0={1\over 2}f_0.\eqno(2.7)$$
To compute the $1/m_Q$ corrections to $f_0$ and $g_0$, we
insert operators $O_1$ and $O_2$,
defined in Eqs. (1.15c) and (1.15d), into the
relevant decay amplitudes:
$$\eqalign{& \Delta M\equiv \Delta M[P^*(v,\varepsilon)\rightarrow P(v)+\pi^a
(q)] \cr     &={1\over f_{\pi}}q^{\mu}\bra {P(v)} iT\int d^4x [O_1(x)+O_2(x)]
\A_{\mu}^a(0)\ket   {P^*(v,\varepsilon)},\cr}\eqno(2.8)$$
$$\eqalign{& \Delta M'\equiv\Delta M[P^*(v,\varepsilon)\rightarrow P^*(v,
\varepsilon^{\prime}) +\pi^a(q)] \cr
&={1\over f_{\pi}}q^{\mu}\bra {P^*(v,\varepsilon^{\prime})} iT\int d^4x
[O_1(x)+O_2(x)]\A_{\mu}^a(0)\ket  {P^*(v,\varepsilon)}.\cr}\eqno(2.9)$$
To determine the general Lorentz structure of Eqs.(2.8) and (2.9), we
recall that the interpolating fields for pseudoscalar and vector mesons are
given by [24]:
$$P_i(v)=\bar{q}\gamma_5h_v^i\sqrt{M_P},\eqno(2.10a)$$
$$P^*_i(v,\varepsilon)=\bar{q}
\varepsilon \!\!\!/ h_v^i\sqrt{M_{P^*}}.\eqno(2.10b)$$
Since we will keep only leading terms in the $1/M_H$ expansion, we can simply
neglect the $1/M_H$ corrections needed for reparametrization invariance.
For the same reason,
we can also neglect residual momenta $k$ and $k'$  in Eqs.
(2.8), (2.9) and (2.10). Furthermore, we shall treat contributions
from $O_1$ and $O_2$ separately. Since $O_1$ preserves
heavy quark spin symmetry, its contributions to both amplitudes must be of
the following forms:
$$\Delta M_1={ ag_s\over f_{\pi}m_Q}\sqrt{M_PM_{P^*}}(\varepsilon
\cdot q)u(P^*)^*{\tau^a\over 2}u(P),\eqno(2.11a)$$
$$\Delta M_1'=-{ ag_s\over f_{\pi}m_Q}M_{P^*}u(P^*)^*{\tau^a\over 2}u(P^
{\prime*})i\epsilon_{\mu\nu\lambda\kappa}
q^{\mu}\varepsilon^{\prime\nu}v^{\lambda}\varepsilon^{\kappa},
\eqno(2.11b)$$
where $u(P^*)$, $u(P)$ and $u(P^{\prime *})$ are isospin wave functions of the
 heavy mesons, and $ a$ is a constant independent of heavy quark masses.

The contributions from $O_2$ are given by
$$\Delta M_2=-{g_sq^{\mu}\over 4m_Qf_{\pi}} \bra {P(v)} iT\int d^4x\,\bar{h}_v
\sigma^{\alpha\beta}G_{\alpha\beta}h_v
(x)\A_{\mu}^a(0)\ket {P^*(v,\varepsilon)},\eqno(2.12a)$$
$$\Delta M_2'=-{g_sq^{\mu}\over 4m_Qf_{\pi}} \bra {P^*(v,\varepsilon^
{\prime})} iT\int d^4x \,\bar{h}_v\sigma^{\alpha\beta}
G_{\alpha\beta}h_v(x) \A_{\mu}^a(0)\ket {P^*(v,\varepsilon)}.
\eqno(2.12b)$$
To evaluate $\Delta M_2$ and $\Delta M_2'$, we make use of Eqs.
(2.10a) and (2.10b) to obtain:
$$\eqalign{\Delta M_2 & =-{g_sq^{\mu}\sqrt{M_PM_{P^*}}\over 4m_Q f_{\pi}}
\bra {0}iT\int d^4x\,\bar{q}_{v}\gamma_5 h_{v} \bar{h}_v\sigma^{\alpha\beta}G_
{\alpha\beta}h_v(x)\A_{\mu}^a(0)\bar{h}_v \varepsilon \!\!\!/\ q_v\ket {0}\cr
&={g_sq^{\mu}\sqrt{M_PM_{P^*}}\over 4m_Q f_{\pi}}
\tr\left(\gamma_5{1+v\!\!\!/\ \over 2}\sigma^{\alpha\beta}{1+v\!\!\!/\ \over
2}\varepsilon \!\!\!/\ \bra {0}iT\int d^4x\,
q_v G_{\alpha\beta}\A^a_\mu\bar{q}_v\ket {0}\right),\cr}
\eqno(2.13a)$$
$$\eqalign{ \Delta M_2' &=-{g_sq^{\mu}\sqrt{M_{P^*}M_{P^*}}\over 4m_Q f_{\pi}}
\bra {0}iT\int d^4x\,\bar{q}_{v}{\varepsilon \!\!\!/\ }^{\prime}h_{v}
\bar{h}_v \sigma^{\alpha\beta}G_{\alpha\beta}h_v(x)\A_{\mu}^a(0)
\bar{h}_v\varepsilon \!\!\!/\ q_v\ket {0}\cr
&={g_sq^{\mu}M_{P^*}\over 4m_Q f_{\pi}}
\tr\left({\varepsilon \!\!\!/\ }^{\prime} {1+v\!\!\!/\ \over 2}\sigma^{\alpha
\beta} {1+v\!\!\!/\ \over 2}\varepsilon \!\!\!/\ \bra {0}
iT\int d^4x \,q_v G_{\alpha\beta}\A_{\mu}^a \bar{q}_v\ket {0}\right).\cr}
\eqno(2.13b)$$
Both of Eqs. (2.13a) and (2.13b) contain a matrix element which describes the
dynamics of light constituents:
$$\eqalign{M_{\alpha\beta\mu} &=\bra{0}iT\int d^4x q_v G_{\alpha\beta}\A_{\mu}
^a\bar{q}_v\ket{0}\cr
&=u(P^*)^*{\tau^a\over 2}u(P)[\epsilon_{\alpha\beta\mu\nu}(bv^{\nu}+c\gamma^{
\nu}+d\gamma^{\nu}v \!\!\!/\ ) +e_1\gamma_5\sigma_{\alpha\beta}v_{\mu}\cr
&+ie_2
\gamma_5(g_{\alpha\mu}\gamma_\beta-g_{\beta\mu}\gamma_\alpha)],  \cr}
\eqno(2.14)$$
where $b$, $c$, $d$, $e_1$ and $e_2$ are constants independent of heavy quark
masses. Note that we have suppressed the $q$-dependent terms since they
correspond to higher-dimensional terms in chiral expansion.
The right hand side of Eq.(2.14) is the most general expression consistent
with the symmetry properties of its left hand side.
Substituting Eq.(2.14) into Eq.(2.13) yields
$$\Delta M_2 =-{g_s\sqrt{M_PM_{P^*}}\over 4m_Q f_{\pi}}(\varepsilon
\cdot q) \times
4u(P^*)^*{\tau^a\over 2}u(P)(b-c+d+2e_2), \eqno(2.15a)$$
$$\Delta M_2' =-{g_s M_{P^*}\over 4m_Q f_{\pi}}  \times
4iu(P^*)^*{\tau^a\over 2}u(P^{\prime *})(b-c+d)\epsilon_{\mu\nu\alpha\beta}
q^\mu\varepsilon^{\prime\nu}v^{\alpha}\varepsilon^{\beta}. \eqno(2.15b)$$
Since $\Delta M$ and $\Delta M'$ can also be computed through the
chiral Lagrangian by expanding
$${\cal A_{\mu}}=-{1\over f_{\pi}}\partial_{\mu}({1\over 2}\tau^a\pi^a)+
\cdots,\eqno(2.16)$$
hence the $\Delta M$ and $\Delta M'$ given by (2.11) and (2.15) imply
$$f_c={2g_s\over \Lambda}[a-(b-c+d+2e_2)],\eqno(2.17a)$$
and
$$g_c= {g_s\over \Lambda}[a+(b-c+d)].\eqno(2.17b)$$
It is clear that
$g_c\not={1\over 2}f_c$ in general.

   We note that the same combination $b-c+d$ appears in both $f_c$ and $g_c$.
It means that the corrections $f_c$ and $g_c$ are characterized by {\it three}
parameters $a,~b-c+d$ and $e_2$! We will now show that $e_2$ is zero, so
actually there are only two unknowns to describe the two coupling constants
$f_c$ and $g_c$. The two processes $P^*\ri P+\pi$ and $P\ri P^*+\pi$ are
related by charge conjugation, and the appropriate coupling constants are $f$
and $f^*$,
respectively. In Eqs.(1.1) and (1.8), it is implicitly assumed that $f$ is
real; this can always be accomplished with a judicious choice of phases for
the field operators of the heavy mesons. We
will assume that this is done. Let us denote
$$\Delta M''\equiv\Delta M[P(v)\ri P^*(v,\varepsilon)+\pi^a(q)],\eqno(2.18)$$
which can be computed by the same procedure for computing $\Delta M$. We find
that $\Delta M''$ also depends on $M_{\alpha\beta\mu}$ given by (2.14).
Indeed, we obtain
$$\Delta M'' =-{g_s\sqrt{M_PM_{P^*}}\over 4m_Q f_{\pi}}(\varepsilon
\cdot q)   \times
4u(P^*)^*{\tau^a\over 2}u(P)(b-c+d-2e_2), \eqno(2.19)$$
which gives
$$f_c^*=\,{2g_s\over\Lambda}[a-(b-c+d-2e_2)].\eqno(2.20)$$
We now demand $f_c=f_c^*$. A comparsion of (2.17) and (2.20) yields $e_2=0$.
Finally,
$$f_c={2g_s\over \Lambda}(a-b'),\eqno(2.21a)$$
and
$$g_c= {g_s\over \Lambda}(a+b'),\eqno(2.21b)$$
where $b'=b-c+d$.

   To discuss $1/m_Q$ corrections to the coupling constants in $\L^{(2)}_{
v,PP^*}$, we shall treat
the heavy-quark and light-quark electromagnetic currents separately. In
the case of the heavy-quark electromagnetic current, the relevant coefficients
$d'$ and $d''\gamma$ are both of order $1/m_Q$ because they arise from the
magnetic moment of the heavy quark. As pointed out by us [17] and by others
[15,16], these couplings are rigorously determined
by the heavy quark effective theory. For completeness, we shall
reproduce the results here. First of all, the heavy-quark electromagnetic
current in the effective theory to order $1/m_Q$ can be written as [9]
$$\eqalign{J^{em}_{\mu}&=\bar{h}_{v'}\gamma_{\mu}h_v-{i\over 2m_Q}
\bar{h}_{v'}(\overleftarrow{D\!\!\!\!/\ }\gamma_{\mu}-\gamma_{\mu}
{D\!\!\!\!/\ })h_v\cr &=\bar{h}_{v'}\gamma_{\mu}h_v-{i\over 2m_Q}
\bar{h}_{v'}(\overleftarrow{D\!\!\!\!/\ }\gamma_{\mu}+\gamma_{\mu}
\overleftarrow{D\!\!\!\!/\ })h_v
+{i\over 2m_Q}\partial^{\nu}(\bar{h}_{v'}\gamma_{\mu}\gamma_{\nu}h_v).
\cr}\eqno(2.22)$$
Using the identity
$$\gamma_{\mu}\gamma_{\nu}=g_{\mu\nu}-i\sigma_{\mu\nu},\eqno(2.23)$$
the Gordon decomposition
$$\bar{h}_{v'}\gamma_{\mu}h_v={1\over 2}\bar{h}_{v'}(v'+v)_{\mu}
h_v+{1\over 2}i\bar{h}_{v'}\sigma_{\mu\nu}(v'-v)^{\nu}h_v,\eqno(2.24)$$
and the identity
$$\bra {H_f(v')}\partial_{\mu}(\bar{h}_{v'}\Gamma h_v)\ket {H_i(v)}
   =i\bar{\Lambda}(v'-v)_{\mu}\bra {H_f(v')}\bar{h}_{v'}\Gamma h_v\ket
{H_i(v)},  \eqno(2.25)$$
with $H_f$ and $H_i$ being generic hadronic states and $\bar{\Lambda}=
M_{H_f}-m_Q=M_{H_i}-m_Q$,
we finally arrive at
$$J_{\mu}^{em}\doteq {1\over 2}\bar{h}_{v'}(v'+v)_{\mu}h_v-{i\over 2m_Q}k^
{\nu}\bar{h}_{v'}\sigma_{\mu\nu}h_v+O_{\mu}^{vv'},\eqno(2.26)$$
where $k^{\nu}=-M_H(v'-v)^{\nu}$, and
$$O_{\mu}^{vv'}=-{i\over m_Q}\bar{h}_{v'}\overleftarrow{D}_{\mu}
h_v-{\bar{\Lambda}\over 2m_Q}(v'-v)^{\nu}\bar{h}_{v'}\gamma_\mu\gamma_\nu
h_v.\eqno(2.27)$$
In Eq.(2.26), we have used the notation
$\doteq$ to remind the reader that such relation holds only after taking the
matrix element.
The contribution due to $O_{\mu}^{vv'}$ is negligible since it cannot change
the normalization of $J_{\mu}^{em}$ at $v=v'$, and its contribution
to the anomalous
magnetic coupling is necessarily of  order ${1\over m_Q^2}$.
The first term in Eq. (2.26) corresponds to the convection current of the
heavy quark, which has already been taken into account in $\L^{(1)}_{v,PP^*}$.
The second term, which is of order $1/m_Q$, will contribute to the
coefficients $d'$ and $d''\gamma$ in $\L^{(2)}_{v,PP^*}$. There is another
source of ${1\over m_Q}$ corrections which arises when one evaluates the
time-ordered products of ${1
\over 2}\bar{h}_{v'}(v'+v)_{\mu}h_v$
with the symmetry-breaking operators $O_1$
 and $O_2$. However, these contributions vanish at $v=v'$ since
the normalization of the vector current is already fixed at the leading order.
Consequently, to order $1/m_Q$, the parameters
$d'$ and $d''\gamma$ are solely induced by the
second term on the r.h.s. of Eq. (2.26):
$$F_{\mu}^Q=\bra {P^*(v',\varepsilon^{\prime})} {ie{\cal Q'}\over 2m_Q}
k^{\nu}\bar{h}_{v'}\sigma_{\mu\nu}h_v\ket {P^*(v,\varepsilon)}|_{v=v'},
\eqno(2.28a)$$
$${\tilde F}_{\mu}^Q=\bra {P(v')} {ie{\cal Q'}\over 2m_Q}
k^{\nu}\bar{h}_{v'}\sigma_{\mu\nu}h_v\ket {P^*(v,\varepsilon)}|_{v=v'}.
\eqno(2.28b)$$
The evaluation of $F_{\mu}^Q$ and ${\tilde F}_{\mu}^Q$ is straightforward
with the aid of Eqs. (2.10a) and (2.10b). Taking $F_{\mu}^Q$ as an example,
we convert the matrix element in Eq. (2.28a) into:
$$\eqalign{F_{\mu}^Q&={ie{\cal Q'}\over 2m_Q}
k^{\nu}M_{P^*}\,\bra {0}\bar{q}_{v'}{\varepsilon\!\!\!/\ }^{\prime}h_{v'}
\bar{h}_{v'}\sigma_{\mu
\nu}h_v \bar{h}_v\varepsilon\!\!\!/\ q_v \ket {0}|_{v=v'}\cr
&=-{ie{\cal Q'}\over 2m_Q}
k^{\nu}M_{P^*}\,\tr\left({\varepsilon\!\!\!/\ }^{\prime} {1+v\!\!\!/\ \over 2}
\sigma_{\mu\nu}{1+v\!\!\!/\ \over 2}\varepsilon\!\!\!/\ \bra {0}q_v\bar{q}_
{v'}\ket {0}|_{v=v'}\right),\cr}\eqno(2.29)$$
where [2]
$$\bra {0}q_v\bar{q}_{v'}\ket {0}|_{v=v'}=\xi(v\cdot v'=1)=1.\eqno(2.30)$$
Working out the trace, we obtain
$$F_{\mu}^Q=-{e{\cal Q'}M_{P^*}\over m_Q}
(\varepsilon_{\mu}\varepsilon^{\prime}
\cdot k-{\varepsilon^{\prime}}_{\mu} \varepsilon \cdot k).\eqno(2.31a)$$
Similarly, we have
$${\tilde F}_{\mu}^Q=-{ie{\cal Q'}\sqrt{M_PM_{P^*}}\over m_Q}
\epsilon_{\mu\nu\alpha\beta}
k^{\nu}\varepsilon^{\alpha}v^{\beta}.\eqno(2.31b)$$
Comparing Eq.(2.31) with Eq.(2.3), we obtain
$$d'=-{e\over 2m_Q}, \ \ \ \  d''\gamma={e\over m_Q}.\eqno(2.32)$$

To determine $d$ and $d''$, we need to consider the form factors induced
by the light-quark electromagnetic current.
To order $1/m_Q$, we have
$$d=d_0+d_c{\Lambda\over 2m_Q},\eqno(2.33a)$$
$$d''=d''_0+d''_c{\Lambda\over 2m_Q}.\eqno(2.33b)$$
As discussed in Ref.[17], one can apply the heavy-quark spin symmetry
to obtain
$$d_0=-{1\over 2}d''_0.\eqno(2.34)$$
To see whether $d_c$ and $d''_c$ obey the same relation,
we evaluate the following magnetic form factors induced by the light quark
electromagnetic current $j_{\mu}^{em}\equiv e\bar{q}{\cal Q}\gamma_{\mu}q$:
$$F_{\mu} =\bra {P^*(v,\varepsilon^{\prime})}
iT\int d^4x [O_1(x)+O_2(x)]j_{\mu}
^{em}(0)\ket {P^*(v,\varepsilon)}_m,
\eqno(2.35a)$$
$${\tilde F}_{\mu}=\bra {P(v)} iT\int d^4x[O_1(x)+O_2(x)]j_{\mu}^{em}(0)\ket
{P^*(v,\varepsilon)}_m,\eqno(2.35b)$$
where the subscript $m$ indicates the fact that we keep only the
magnetic interactions. To evaluate $F_{\mu}$ and ${\tilde F}_{\mu}$, we
again employ the technique of interpolating fields [24] to obtain
$$F_{\mu}=\bra {0}iT\int d^4x\,\bar{q}_v{\varepsilon\!\!\!/\ }^{\prime}h_{v}
[O_1(x)+O_2(x)]j_{\mu}^{em}(0)\bar{h}_v\varepsilon\!\!\!/\
q_v\ket {0}_m,\eqno(2.36a)$$
and
$${\tilde F}_{\mu}=\bra {0}iT\int d^4x\,\bar{q}_v\gamma_5 h_{v}
[O_1(x)+O_2(x)]j_{\mu}^{em}(0)\bar{h}_v\varepsilon\!\!\!/\
q_v\ket {0}_m.\eqno(2.36b)$$
For convenience, we shall treat contributions by $O_1(x)$ and $O_2(x)$
separately. Their contributions are denoted by $F_{\mu}^1\,$(${\tilde
F}_{\mu}^1$) and $F_{\mu}^2\,$(${\tilde F}_{\mu}^2$) respectively.
As $O_1(x)$ preserves the heavy-quark spin symmetry, $F_{\mu}^1$ is
related to ${\tilde F}_{\mu}^1$ in such a way that
$$F_{\mu}^1={2a_1g_s\over m_Q}M_{P^*}(\varepsilon_{\mu}\varepsilon^{\prime}
\cdot k-{\varepsilon^{\prime}}_{\mu} \varepsilon \cdot k),\eqno(2.37a)$$
and
$${\tilde F}_{\mu}^1={2ia_1g_s\over m_Q} \sqrt{M_PM_{P^*}}\epsilon_{\mu\nu
\alpha\beta}k^{\nu}v^{\alpha}\varepsilon^{\beta},\eqno(2.37b)$$
with $a_1$ being a constant independent of the heavy quark mass.
For simplicity, we have set the charge matrix ${\cal Q}={\bf 1}$ and
suppressed the flavor quantum numbers while obtaining Eqs.(2.35) and (2.37).
To compute $F_{\mu}^2$ and ${\tilde F}_{\mu}^2$, we apply Eqs.
(2.10a) and (2.10b) to obtain
$$\eqalign{F_{\mu}^2 &=-{g_sM_{P^*}\over 4m_Q}\bra {0}iT\int d^4x\,\bar{q}_{v}
{\varepsilon\!\!\!/\ }^{\prime}h_{v}
\bar{h}_v\sigma^{\alpha\beta}G_{\alpha\beta}h_v(x)j_{\mu}^{em}(0)
\bar{h}_v\varepsilon\!\!\!/\ q_v\ket {0}_m\cr
&= {g_sM_{P^*}\over 4m_Q}\tr\left( {\varepsilon \!\!\!/\ }^{\prime}
{1+v\!\!\!/\ \over 2}\sigma^{\alpha\beta}{1 +v\!\!\!/\ \over 2}\varepsilon
\!\!\!/\bra {0}iT\int d^4x \,q_{v}G_{\alpha\beta}j_{\mu}^{em}\bar{q}_v\ket {0}
\right)_m,\cr} \eqno(2.38a)$$
$$\eqalign{{\tilde F}_{\mu}^2 &=-{g_s\sqrt{M_PM_{P^*}}\over 4m_Q}
\bra {0}iT\int d^4x\,\bar{q}_v\gamma_5 h_{v}
\bar{h}_v\sigma^{\alpha\beta}G_{\alpha\beta}h_v(x)j_{\mu}^{em}(0)
\bar{h}_v\varepsilon\!\!\!/\ q_v\ket {0}_m\cr
&= {g_s\sqrt{M_PM_{P^*}}\over 4m_Q}\tr\left(\gamma_5 {1+
v\!\!\!/\ \over 2}\sigma^{\alpha\beta}{1+v\!\!\!/\ \over 2}
\varepsilon\!\!\!/ \bra {0}iT\int d^4x \,q_{v}G_{\alpha\beta}j_{\mu}^{em}
\bar{q}_v\ket {0}\right)_m.\cr}
\eqno(2.38b)$$
The non-perturbative dynamics of the light constituents can be parametrized as
$$\eqalign{M'_{\alpha\beta\mu} &=\bra {0}iT\int d^4x \,q_{v}G_{\alpha\beta}j_{
\mu}^{em}\bar{q}_v \ket {0}\cr
&=-ib_1(g_{\alpha\mu}k_{\beta}-g_{\beta\mu}k_{\alpha}), \cr}\eqno(2.39)$$
where $k$ is the outgoing photon momentum and $b_1$ is a constant independent
of the heavy quark mass. In Eq.(2.39), we kept only structures linear in $k$
relevant to magnetic interactions. With this to be understood, Eq.(2.39) then
represents the most general Lorentz structure for $M'_{\alpha\beta\mu}$ which
is consistent with gauge invariance, parity conservation, and the constraint
$M'_{\alpha\beta\mu}=-M'_{\beta\alpha\mu}$. Substituting Eq.(2.39) into Eqs.
(2.38a) and (2.38b), we obtain
$$F_{\mu}^2={-g_sM_{P^*}\over m_Q}b_1(\varepsilon_{\mu}\varepsilon^{
\prime}\cdot k-{\varepsilon^{\prime}}_{\mu} \varepsilon \cdot k),
\eqno(2.40a)$$
$${\tilde F}_{\mu}^2={ig_s\sqrt{M_PM_{P^*}}\over m_Q}b_1
\epsilon_{\mu
\nu\alpha\beta}k^{\nu}v^{\alpha}\varepsilon^{\beta}. \eqno(2.40b)$$
Since the results in Eqs.(2.40a) and (2.40b) can also be obtained from the
Lagrangian $\L^{(2)}_{v,PP^*}$, we can hence make the following
identifications:
$$d_c=\,{g_s\over \Lambda}(2a_1+b_1),\eqno(2.41a)$$
$$d''_c=\,{2g_s\over \Lambda}(-2a_1+b_1).\eqno(2.41b)$$
It is clear that $d_c\not=-{1\over 2}d''_c$ in general.

Eqs.(2.21) and (2.41) are the main results in this section.

\endpage

\noindent{\bf 3.~~$1/m_Q$ Corrections to the Dynamics of Heavy Baryons}
\vskip 0.3 cm

In this section we
study the $1/m_Q $ corrections to the coupling constants appearing in the
heavy-baryon chiral Lagrangian ${\cal L}_B^{(1)}$ and ${\cal L}_B^{(2)}$
given by Eqs. (3.8) and (3.9) respectively in Ref.[17].
In terms of velocity-dependent fields,
$$\eqalign{{\cal L}_{v,B}^{(1)}&={1\over 2} \tr[\bar{B}_{\bar 3}(v)(iv\cdot D)
B_{\bar 3}(v)]+\tr[\bar{B}_{6}(v)(iv\cdot D)B_{6}(v)]\cr
&-\tr [\bar{B}_6^{*\mu}(v)(iv\cdot D)B_{6\mu}^{*}(v)]\cr
&+g_1\tr[\bar{B}_6(v)\gamma_{\mu}\gamma_5{\cal A^{\mu}}B_6(v)]
+g_2\tr[\bar{B}_6(v)\gamma_{\mu}\gamma_5{\cal A^{\mu}}B_{\bar{3}}(v)]+h.c.\cr
&+g_3\tr[\bar{B}_{6\mu}^*(v){\cal A^{\mu}}B_6(v)]+h.c.
+g_4\tr[\bar{B}_6^{*\mu}(v){\cal A_{\mu}}B_{\bar 3}(v)]+h.c.\cr
&+g_5\tr[\bar{B}_6^{*\nu}(v)\gamma_{\mu}\gamma_5{\cal A^{\mu}}
B_{6\nu}^*(v)] +g_6\tr[\bar{B}_{\bar{3}}(v)\gamma_{\mu}\gamma_5
{\cal A^{\mu}}B_{\bar{3}}(v)],\cr}\eqno(3.1)$$
with
$$D_{\mu}B(v)=\partial_{\mu}B(v)+{\cal V_{\mu}}B(v)+B(v){\cal V_{\mu}}^T
+ie{\cal Q^{\prime}}A_{\mu}B(v)+ieA_{\mu}\{ {\cal Q},~B(v)\},\eqno(3.2)$$
where as before $\Q={\rm diag}({2\over 3},-{1\over 3},-{1\over 3})$ is the
charge matrix of light quarks and $\Q'$ is the charge of the heavy quark $Q$.
Note that we have omitted those terms which are induced by mass differences
between various baryons.

It is well known that baryons do not behave much like Dirac point particles.
As a result, they can have large
anomalous magnetic moments. The most general gauge invariant Lagrangian for
magnetic transitions of heavy baryons is given by
$$\eqalign{{\cal L}_{v,B}^{(2)}&=a_1\tr[\bar{B}_6(v){\cal Q}\sigma\cdot F B_6
(v)]+a_1^{\prime}\tr[\bar{B}_6(v){\cal Q^{\prime}}\sigma\cdot F B_6(v)]\cr
&+a_2\tr[\bar{B}_6(v){\cal Q}\sigma\cdot F B_{\bar 3}(v)]+h.c.
+a_2^{\prime}\tr[\bar{B}_6(v){\cal Q^{\prime}}\sigma\cdot F B_{\bar 3}(v)]+
h.c.\cr
&+a_3\tr[\epsilon_{\mu\nu\lambda\kappa}\bar{B}_6^{*\mu}(v)
{\cal Q}\gamma^{\nu}F^{\lambda\kappa}B_6(v)]+h.c.\cr
&+a_3^{\prime}\tr[\epsilon_{\mu\nu\lambda\kappa}\bar{B}_6^{*\mu}(v)
{\cal Q^{\prime}}\gamma^{\nu}F^{\lambda\kappa}B_6(v)]+h.c.\cr
&+a_4\tr[\epsilon_{\mu\nu\lambda\kappa}\bar{B}_6^{*\mu}(v)
{\cal Q}\gamma^{\nu}F^{\lambda\kappa}B_{\bar 3}(v)]+h.c.\cr
&+a_4^{\prime}\tr[\epsilon_{\mu\nu\lambda\kappa}\bar{B}_6^{*\mu}(v)
{\cal Q^{\prime}}\gamma^{\nu}F^{\lambda\kappa}B_{\bar 3}(v)]+h.c.\cr
&+a_5\tr[\bar{B}_6^{*\mu}(v){\cal Q}\sigma\cdot FB_{6\mu}^*(v)]
+a_5^{\prime}\tr[\bar{B}_6^{*\mu}(v){\cal Q^{\prime}}\sigma\cdot FB_{6\mu}^*
(v)]\cr
&+a_6\tr[\bar{B}_{\bar 3}(v){\cal Q}\sigma\cdot F B_{\bar 3}(v)]
+a_6^{\prime}\tr[\bar{B}_{\bar 3}(v){\cal Q^{\prime}}\sigma\cdot F B_{\bar 3}
(v)].\cr}\eqno(3.3)$$
The Lagrangian ${\cal L}_{v,B}^{(2)}$
is also the most general chiral-invariant
one provided that one makes the replacement
$${\cal Q}\rightarrow {1\over 2}(\xi^{\dagger}{\cal Q}\xi+\xi {\cal Q}
\xi^{\dagger}), \ \ \ {\cal Q^{\prime}}\rightarrow {\cal Q^{\prime}}.
\eqno(3.4)$$
Note that, contrary to Eq. (3.9) of Ref.[17], we do not need to subtract from
Eq.(3.3) the Dirac magnetic moments of the heavy baryons, because ${\cal L}_{
v,B}^{(1)}$ is now expressed in terms of velocity dependent fields and
consequently contains no Dirac magnetic moments to the lowest order. The
magnetic couplings $a_i$ are induced by light-quark electromagnetic currents
whereas $a^{\prime}_i$ are induced by heavy-quark ones and they are of order
$1/m_Q$.

To incorporate $1/m_Q$ corrections, we expand the coupling constants in ${\cal
L}_{v,B}^{(1)}$ and ${\cal L}_{v,B}^{(2)}$ as follows:
$$g_i=g_i^0+g_i^c {\Lambda\over 2m_Q},\eqno(3.5a)$$
$$a_i=a_i^0+a_i^c {\Lambda\over 2m_Q}. \eqno(3.5b)$$
Let us first focus on the coupling constants $g_i$'s in ${\cal L}_{v,B}^{
(1)}$. The relations among the leading terms $g_i^0$ are governed by HQS [11].
 They have been derived by evaluating the decay amplitudes $B_{\bar 3}
\rightarrow B_{\bar 3}+\pi$, $B_6(B_6^*)\rightarrow B_{\bar 3}+\pi$, and
$B_6(B_6^*)\rightarrow B_6(B_6^*)+\pi$. The results can be summarized
as follows:
$$g_3^0={\sqrt 3\over 2}g_1^0, \ \ \ g_5^0=-{3\over 2}g_1^0,\eqno(3.6a)$$
$$g_4^0=-{\sqrt 3}g_2^0,\eqno(3.6b)$$
$$g_6^0=0.\eqno(3.6c)$$
The result $g_6^0=0$ follows from the fact that, in the heavy quark spin
symmetry limit, the
strong transition between antitriplet baryons is forbidden by parity
conservation.
To relate the sub-leading coefficients $g_i^c$'s, we insert the operators
$O_1$ and $O_2$ of Eqs. (1.15c) and (1.15d) into the relevant matrix elements.
First of all, the sub-leading amplitude for $B_{\bar3}\rightarrow
B_{\bar 3}+\pi$ is given by
$$\eqalign{&\Delta M_{{\bar 3}}[B_{\bar 3}(v,s)\rightarrow B_{\bar 3}(v,s')+
\pi^a(q)]
\cr  &={1\over f_{\pi}}q^{\mu}\bra {B_{\bar 3}(v,s')}iT\int d^4x [O_1(x)+O_2
(x)]\A_{\mu}^a(0)\ket {B_{\bar 3}(v,s)}.\cr}\eqno(3.7)$$
To determine the general Lorentz structure of $M_{{\bar 3}}$, we employ
the following interpolating field for antitriplet baryons [25]:
$$B_{\bar 3}(v,s)={\bar u}(v,s)\phi_vh_v,\eqno(3.8)$$
where $\phi_v$ is a Lorentz scalar.
One can easily show that $O_1$ does not contribute to $M_{{\bar 3}}$. Denoting
 the $O_1$'s contribution as $\Delta M_{{\bar 3}}^1$, we then have
$$\eqalign{& \Delta M_{{\bar 3}}^1[B_{\bar 3}(v,s)\rightarrow B_{\bar 3}(v,
s')+\pi^a(q)]\cr    &={1\over 2f_{\pi}m_Q}q^{\mu}\bra {B_{\bar 3}(v,s')}iT\int
 d^4x\,\bar{h}_v
(iD)^2h_v(x)\A_{\mu}^a(0)\ket {B_{\bar 3}(v,s)}.\cr}\eqno(3.9)$$
By Eq.(3.8) we may rewrite Eq.(3.9) as
$$\Delta M_{{\bar 3}}^1={1 \over 2f_{\pi}m_Q}q^{\mu}\bar{u}(v,s')u(v,s)
\bra {0}iT\int d^4x\,\phi_{v}(iD)^2 \A_{\mu}^a\phi_v^{\dagger}\ket {0}.
\eqno(3.10)$$
Since we cannot construct an axial vector out of $v$ and $q$, we conclude that
$$\bra {0}iT\int d^4x\, \phi_{v}(iD)^2\A_{\mu}^a\phi_v^{\dagger}\ket {0}=0,
\eqno(3.11)$$
and hence $\Delta M_{\bar{3}}^1=0$.
The situation is different in the case of $O_2$-insertion.
The amplitude $\Delta M_{{\bar 3}}^2$ is given by
$$\Delta M_{{\bar 3}}^2=-{g_sq^{\mu}\over 4m_Qf\pi}
\bra {B_{\bar 3}(v,s')}iT\int d^4x\,\bar{h}_v\sigma^{\alpha\beta}G_{
\alpha\beta}h_v(x)\A_{\mu}^a(0)\ket {B_{\bar 3}(v,s)}.\eqno(3.12)$$
Applying Eq.(3.8) gives
$$\Delta M_{{\bar 3}}^2=-{g_sq^{\mu}\over 4m_Qf_\pi}
\bar{u}(v,s'){1+v\!\!\!/\ \over 2}\sigma^{\alpha\beta}{1+v\!\!\!/\ \over 2}u(
v,s)\bra {0}iT\int d^4x\,\phi_vG_{\alpha\beta}\phi_v^{\dagger}\A_{\mu}^a\ket
{0}.\eqno(3.13)$$
Since the diquark field $\phi_v$ is a Lorentz scalar, we may parametrize the
matrix element of light constituents as
$$\bra {0}iT\int d^4x\,\phi_vG_{\alpha\beta}\phi_v^{\dagger}\A_{\mu}^a\ket {0}
=-r\epsilon_{\alpha\beta\mu\nu}v^{\nu},\eqno(3.14)$$
where $r$ is a constant independent of the heavy quark mass. We have also
neglected the flavor wave functions of incoming and outgoing baryons as they
are irrelevant to our discussion. Substituting
Eq.(3.14) into Eq.(3.13) yields
$$\Delta M_{{\bar 3}}^2={rg_s\over 2m_Qf_{\pi}}\bar{u}(v,s')q\!\!\!/
\gamma_5u(v,s).\eqno(3.15)$$
Comparing this result with Eqs.(3.1) and (3.5), we find
$$g_6^c={rg_s\over \Lambda},\eqno(3.16)$$
which is non-vanishing in general. This shows that
the decay $B_{\bar 3}\rightarrow B_{\bar 3}+\pi$, while forbidden
in the infinitely heavy quark limit, is allowed in the sub-leading order.
As a similar conclusion has also been arrived by Cho [18], we would like to
compare our result with his in some details.

Cho constructed the following operator to describe the decay
$B_{\bar 3}\rightarrow B_{\bar 3}+\pi$ to the order of ${1/m_Q}$:
$$O_{TTA}={i\over m_Q}
\epsilon_{\mu\nu\sigma\lambda}\bar{T}^j(v)\sigma^{\mu\nu}
D^{\sigma}T_i(v){(\A^{\lambda})}^i_j,\eqno(3.17)$$
where
$$T(v)_i=\epsilon_{ijk}(B_{\bar 3}(v))_{jk},\eqno(3.18)$$
apart from an overall normalization. In Eq.(3.18), $(B_{\bar 3})_{jk}$ is the
$jk$ matrix element in the baryon matrix $B_{\bar 3}$ [11].
The operator $O_{TTA}$ is not
reparametrization  invariant itself. Therefore, it should be part of one which
is. We will now show that $O_{TTA}$ is a reparametrization invariant
partner to the interaction term in (3.1) with the coupling constant $g_6$:
$$\L_6=g_6\tr[\bar{B}_{\bar 3}\gamma_{\mu}\gamma_5{\cal A}^{\mu}B_{\bar 3}],
\eqno(3.19)$$
where it is understood
that $B_{\bar{3}}$ and $\bar{B}_{\bar{3}}$ have velocity
$v$. The reparametrization invariant generalization of $\L_6$ is obtained
through the substitution [20]
$$B_{\bar 3}\rightarrow \left(1+{iD\!\!\!\!/\ \over 2M_{\bar 3}}\right)
B_{\bar 3}.\eqno(3.20)$$
The result is
$$\eqalign{ \tilde{\L}_6 & =g_6\tr\l[\bar{B}_{\bar{3}}\Big(1-{i\ol{D\!\!\!\!/
}\over 2M_{\bar 3}}\Big)\gamma_\mu\gamma_5\A^\mu\Big(1+{iD\!\!\!\!/ \over 2M_{
\bar 3}}\Big)B_{\bar{3}}\r]  \cr   & =\,\L_6+\L'_6,  \cr}\eqno(3.21)$$
where $\L_6$ is given by (3.19) and
$$\L'_6=-{ig_6\over 2M_{\bar{3}}}\,\tr\left\{\bar{B}_{\bar{3}}[\ol{D\!\!\!\!/\
}\gamma_\mu\A^\mu+\gamma_\mu\A^\mu D\!\!\!\!/\ ]\gamma_5 B_{\bar{3}}\right\}.
\eqno(3.22)$$
The identity
$$\gamma_\mu\gamma_\nu=g_{\mu\nu}-i\sigma_{\mu\nu},\eqno(3.23)$$
reduces $\L'_6$ to
$$\L'_6 =-{ig_6\over 2M_{\bar{3}}}\tr\Big\{\bar{B}_{\bar{3}}(\ol{D}
_\mu\A^\mu+\A^\mu D_\mu)\gamma_5 B_{\bar{3}} +
i\bar{B}_{\bar{3}}\sigma_{\mu\nu}(\ol{D}{^\nu}\A^\mu-\A^\mu D^\nu)\gamma_5
B_{\bar{3}}\Big\}.  \eqno(3.24)$$
The first term in (3.24) vanishes as a result of the two identities
$$\left(\bar{B}_{\bar{3}}\right)_{ij}(\ol{\partial}_\mu+\partial_\mu)\gamma_5
\left(B_{\bar{3}}\right)_{kl}=\partial_\mu\left[\left(\bar{B}_{\bar{3}}\right)
_{ij}\gamma_5\left(B_{\bar{3}}\right)_{kl}\right],\eqno(3.25a)$$
$$\left(\bar{B}_{\bar{3}}\right)_{ij}\gamma_5\left(B_{\bar{3}}\right)_{kl}=0.
\eqno(3.25b)$$
The second term in (3.24) can be further transformed with the
aid of the identity
$$\epsilon_{\mu\nu\sigma\lambda}\sigma^{\mu\nu}=2i\sigma_{\sigma\lambda}
\gamma_5.\eqno(3.26)$$
Finally, we obtain
$$\L'_6=\,{ig_6\over 4M_{\bar{3}}}\epsilon_{\mu\nu\sigma\lambda}\tr\left[\bar{
B}_{\bar{3}}\sigma^{\mu\nu}(\ol{D}{^\sigma}\A^\lambda-\A^\lambda D^\sigma)B_{
\bar{3}}\right],\eqno(3.27)$$
which aside from the factor $g_6$ is $O_{TTA}$ in a somewhat different
notation. We see that as a reparametrization invariant partner to $\L_6$,
the coupling constant for $O_{TTA}$ is given by $g_6$ which is of order
$O(1/m_Q)$. Since the operator $O_{TTA}$ already contains a factor of $1/m_Q$,
its contribution is smaller by one power of $1/m_Q$ relative to $\L_6$.

Our next task is to relate $g_2^c$ to $g_4^c$. To do this, we evaluate
the amplitudes of $B_6\rightarrow B_{\bar 3} +\pi$ and
$B_6^*\rightarrow B_{\bar 3}+ \pi$. With our previous
notations, the sub-leading contribution is
$$\eqalign{&\Delta M_{6{\bar 3}}[B_6(v,s,\kappa)\rightarrow B_{\bar 3}(v,s')
+\pi^a(q)]  \cr    &={1\over f_{\pi}}q^{\mu}\bra {B_{\bar 3}(v,s')}iT\int d^4x
[O_1(x)+O_2(x)]\A_{\mu}^a(0)\ket {B_6(v,s,\kappa)},\cr}\eqno(3.28)$$
where $\kappa$ is used to specify the spin of sextet baryons:
$\kappa =1$ corresponds to spin-${1\over 2}$ baryons whereas
$\kappa=2$ denotes spin-${3\over 2}$ ones. To evaluate
$\Delta M_{6{\bar 3}}$, we employ the following interpolating fields [17]
$$B_6(v,s,\kappa)=\bar{B}_{\mu}(v,s,\kappa)\phi_v^{\mu}h_v,\eqno(3.29)$$
where $\phi_v^{\mu}$ is an axial vector field. The wave function $\bar{B}_
{\mu}$ is given by
$$\bar{B}_{\mu}(v,s,\kappa=1)={1\over \sqrt 3}\bar{u}(v,s)\gamma_5 (\gamma_{
\mu}+v_{\mu}),\eqno(3.30a)$$
$$\bar{B}_{\mu}(v,s,\kappa=2)=\bar{u}_{\mu}(v,s), \eqno(3.30b)$$
with $u_{\mu}(v,s)$
and $u(v,s)$ being the Rarita-Schwinger's vector spinor and
usual Dirac spinor respectively.
The contribution from the operator $O_2$ gives
$$\eqalign{\Delta M_{6{\bar 3}}^2&=-{g_sq^{\mu}\over 4m_Qf_{\pi}}
\bra {B_{\bar 3}(v,s')}iT\int d^4x \bar{h}_v\sigma^{\alpha\beta}G_{\alpha
\beta}h_v(x)\A_{\mu}^a(0)\ket {B_6(v,s,\kappa)}\cr
&=-{g_sq^{\mu}\over 4m_Qf_{\pi}}
\bar{u}(v,s'){1+v\!\!\!/\ \over 2}\sigma^{\alpha\beta}{1+v\!\!\!/\ \over 2}B^
{\nu}(v,s,\kappa)\times \cr
&\bra {0}iT\int d^4x\,\phi_vG_{\alpha\beta}\A_{\mu}^a
\phi_{v,\nu}^{\dagger}\ket {0}.\cr}\eqno(3.31)$$
The matrix element for light constituents may be parametrized as
$$\eqalign{M_{\mu\nu\alpha\beta} &=\bra {0}\phi_{v}G_{\alpha\beta}A_{\mu}^a
\phi_{v,\nu}^{\dagger}\ket {0}\cr
 &=r_1(g_{\mu\alpha}v_{\beta}-g_{\mu\beta}v_{\alpha})v_{\nu}+r_2(g_{\nu\alpha}
v_{\beta}-g_{\nu\beta}v_{\alpha})v_{\mu}\cr
&+ir_3(g_{\alpha\mu}g_{\beta\nu}-g_{
\beta\mu}g_{\alpha\nu}), \cr}\eqno(3.32)$$
where $r_1$, $r_2$ and $r_3$ are constants independent of the heavy quark
mass, and the flavor wave functions are neglected for simplicity. This
is the most general Lorentz structure for $M_{\mu\nu\alpha\beta}$
which is antisymmetric in $\alpha$ and $\beta$. With Eq.(3.32)
we can immediately conclude that contributions from $r_1$ and $r_2$ are zero
because of the identity
$$ {1+v\!\!\!/\ \over 2}\sigma^{\alpha\beta}{1+v\!\!\!/\ \over 2}v_{\alpha}=0.
\eqno(3.33)$$
The contribution due to $r_3$ is
$$\Delta M^2_{6\bar{3}}=-i{r_3g_sq^\mu\over 2m_Qf_\pi}\,\bar{u}\sigma_{\mu\nu}
B^\nu(v,s,\kappa).\eqno(3.34)$$
Using (3.30) for the wave function $B^\nu$, we find
$$\Delta M^2_{6\bar{3}}(6^*\ri \bar{3})=-{r_3g_s\over 2m_Qf_\pi}\,
\bar{u}q^\mu u_\mu,\eqno(3.35a)$$
$$\Delta M^2_{6\bar{3}}(6\ri \bar{3})=-{r_3g_s\over \sqrt{3}m_Qf_\pi}\,
\bar{u}_{\bar 3}q\!\!\!/ \gamma_5 u_6,\eqno(3.35b)$$
where we have used the notations $6^*$ and 6 to denote a spin ${3\over 2}$
and spin ${1\over 2}$ baryon in the sextet, respectively. Eq.(3.35) implies
the following corrections to the coupling constants $g_2$ and $g_4$:
$$g_2^c=\,{g_s\over \Lambda}\l(r'-{2\over\sqrt{3}}r_3\r),\eqno(3.36a)$$
$$g_4^c=\,{g_s\over \Lambda}\l(-\sqrt{3}r'-r_3\r),\eqno(3.36b)$$
where we have added the contributions proportional to $r'$ coming from the
operator $O_1$ which preserves the spin symmetry.

Finally, we discuss strong transitions among sextet baryons. The contribution
due to $O_2$ gives
$$\eqalign{\Delta M_{66}^2&=-{g_sq^{\mu}\over 4m_Qf_{\pi}}
\bra {B_6(v,s',\kappa^{\prime})}iT\int d^4x \bar{h}_v\sigma^{\alpha
\beta}G_{\alpha\beta}h_v(x)\A_{\mu}^a(0)\ket {B_6(v,s,\kappa)}\cr
&=-{g_sq^{\mu}\over 4m_Qf_{\pi}}
\bar{B}^{\rho}(v,s',\kappa^{\prime}){1+v\!\!\!/\ \over 2}\sigma^{\alpha\beta}
{1+v\!\!\!/\ \over 2}B^{\nu}(v,s,\kappa)\times\cr
&\bra {0}iT\int d^4x\,\phi_{v,\rho}G_{\alpha\beta}\A_{
\mu}^a\phi_{v,\nu}^{\dagger}\ket {0}.\cr}\eqno(3.37)$$
The matrix elements of the light constituents can be parametrized as
$$\eqalign{M_{\alpha\beta\mu\lambda\kappa} &=\bra {0}iT\int d^4x\,\phi_{v,
\lambda}G_{\alpha\beta}\A_{\mu}^a\phi_{v,\kappa}^{\dagger}\ket {0}  \cr
 &=v_{\mu}\epsilon_{\lambda\alpha\beta\kappa}\,s+\epsilon_{\alpha\beta\mu\nu}
v^\nu g_{\lambda\kappa}s_1+\epsilon_{\alpha\beta\kappa\nu}v^\nu g_{
\lambda\mu}s_2+\epsilon_{\alpha\beta\lambda\nu}v^\nu g_{\mu\kappa}s_3   \cr
&+(g_{\alpha\mu}\epsilon_{\beta\nu\lambda\kappa}-g_{\beta\mu}\epsilon_{\alpha
\nu\lambda\kappa})v^\nu s_4+(g_{\alpha\lambda}\epsilon_{\beta\nu\mu\kappa}-g_
{\beta\lambda}\epsilon_{\alpha\nu\mu\kappa})v^\nu s_5   \cr
&+(g_{\alpha\kappa}\epsilon_{\beta\nu\mu\lambda}-g_{\beta\kappa}\epsilon_{
\alpha\nu\mu\lambda})v^\nu s_6,   \cr}\eqno(3.38)$$
where $s$ and $s_i$ are constants independent of the heavy quark mass, and
the flavor wave functions are again neglected. This is the most general
Lorentz structure for $M_{\alpha\beta\mu\lambda\kappa}$ which conserves
parity, and is antisymmetric with respect to the indices $\alpha$ and $\beta$.
Let us write
$$\Delta M^2_{66}=-{g_sq_\mu\over 4m_Qf_\pi}\,\Delta,\eqno(3.39)$$
with
$$\Delta=\,\bar{B}_\lambda(v,s',\kappa')\sigma_{\alpha\beta}B_\rho(v,s,
\kappa)M^{\alpha\beta\mu\lambda\rho}.\eqno(3.40)$$
We further denote
$$\Delta=\,\Delta_s+\Delta_1+\Delta_2+\cdots+\Delta_6\eqno(3.41)$$
for contributions due to $s,~s_1,~s_2,\cdots s_6$, respectively.

   Using the properties of the wave functions
$$v^\mu B_\mu(v,s,\kappa)=0,\eqno(3.42a)$$
$$(v\!\!\!/\ -1)B_\mu(v,s,\kappa)=0,\eqno(3.42b)$$
$$\gamma^\mu u_\mu(v,s)=0,\eqno(3.42c)$$
the identity (3.26) and
$$i\epsilon^{\mu\nu\lambda\kappa}\gamma_\kappa=(-\gamma^\mu\gamma^\nu\gamma^
\lambda+g^{\mu\nu}\gamma^\lambda-g^{\mu\lambda}\gamma^\nu+g^{\nu\lambda}\gamma
^\mu)\gamma_5,\eqno(3.43)$$
we obtain
$$\Delta_s=0,\eqno(3.44)$$

$$\Delta_1(6^*\ri 6^*)=\,2s_1\bar{u}^\lambda\gamma^\mu\gamma_5u_\lambda,
\eqno(3.45a)$$
$$\Delta_1(6^*\ri 6)=\,{4\over \sqrt{3}}s_1\bar{u}u^\mu,\eqno(3.45b)$$
$$\Delta_1(6\ri 6^*)=\,{4\over \sqrt{3}}s_1\bar{u}^\mu u,\eqno(3.45c)$$
$$\Delta_1(6\ri 6)=\,{2\over 3}s_1\bar{u}\gamma^\mu\gamma_5u,\eqno(3.45d)$$

$$\Delta_2(6^*\ri 6^*)=\Delta_2(6^*\ri 6)=0,\eqno(3.46a)$$
$$\Delta_2(6\ri 6^*)=2\sqrt{3}s_2\bar{u}^\mu u,\eqno(3.46b)$$
$$\Delta_2(6\ri 6)=-2s_2\bar{u}\gamma^\mu\gamma_5 u,\eqno(3.46c)$$

$$\Delta_3(6^*\ri 6^*)=\Delta_3(6\ri 6^*)=0,\eqno(3.47a)$$
$$\Delta_3(6^*\ri 6)=2\sqrt{3}s_3\bar{u}u^\mu,\eqno(3.47b)$$
$$\Delta_3(6\ri 6)=-2s_3\bar{u}\gamma^\mu\gamma_5 u,\eqno(3.47c)$$

$$\Delta_4(6^*\ri 6^*)=\Delta_4(6\ri 6)=0,\eqno(3.48a)$$
$$\Delta_4(6^*\ri 6)=2\sqrt{3}s_4\bar{u}u^\mu,\eqno(3.48b)$$
$$\Delta_4(6\ri 6^*)=-2\sqrt{3}s_4\bar{u}^\mu u,\eqno(3.48c)$$

$$\Delta_5(6^*\ri 6^*)=\,2s_5\bar{u}^\lambda\gamma^\mu\gamma_5u_\lambda,
\eqno(3.49a)$$
$$\Delta_5(6^*\ri 6)=\,{4\over \sqrt{3}}s_5\bar{u}u^\mu,\eqno(3.49b)$$
$$\Delta_5(6\ri 6^*)=\,-{2\over \sqrt{3}}s_5\bar{u}^\mu u,\eqno(3.49c)$$
$$\Delta_5(6\ri 6)=\,{8\over 3}s_5\bar{u}\gamma^\mu\gamma_5u,\eqno(3.49d)$$

$$\Delta_6(6^*\ri 6^*)=\,2s_6\bar{u}^\lambda\gamma^\mu\gamma_5u_\lambda,
\eqno(3.50a)$$
$$\Delta_6(6^*\ri 6)=\,-{2\over \sqrt{3}}s_6\bar{u}u^\mu,\eqno(3.50b)$$
$$\Delta_6(6\ri 6^*)=\,{4\over \sqrt{3}}s_6\bar{u}^\mu u,\eqno(3.50c)$$
$$\Delta_6(6\ri 6)=\,{8\over 3}s_6\bar{u}\gamma^\mu\gamma_5u.\eqno(3.50d)$$
Collecting all the terms, we find
$$\Delta M^2(6^*\ri 6^*)=-{g_s\over 2m_Qf_\pi}(s_1+s_5+s_6)\bar{u}^\lambda
q\!\!\!/ \gamma_5u_\lambda,\eqno(3.51a)$$
$$\Delta M^2(6^*\ri 6)=-{g_s\over 2m_Qf_\pi}\l({2\over\sqrt{3}}s_1+\sqrt{3}s_3
+\sqrt{3}s_4+{2\over\sqrt{3}}s_5-{1\over\sqrt{3}}s_6\r)\bar{u}
q^\mu u_\mu,\eqno(3.51b)$$
$$\Delta M^2(6\ri 6^*)=-{g_s\over 2m_Qf_\pi}\l({2\over\sqrt{3}}s_1+\sqrt{3}s_2
-\sqrt{3}s_4-{1\over\sqrt{3}}s_5+{2\over\sqrt{3}}s_6\r)\bar{u}_\mu
q^\mu u,\eqno(3.51c)$$
$$\Delta M^2(6\ri 6)=-{g_s\over 2m_Qf_\pi}\l({1\over 3}s_1-s_2-s_3+{4\over 3}
s_5+{4\over 3}s_6\r)\bar{u} q\!\!\!/ \gamma_5u,\eqno(3.51d)$$
where we have dropped the subscripts 66 in the corrections to the matrix
elements $\Delta M^2$. When Eq.(3.51) is compared with (3.1), we find
$$g_1^c=-{g_s\over\Lambda}(s'+{1\over 3}s_1-s_2-s_3+{4\over 3}s_5+{4\over 3}
s_6),\eqno(3.52a)$$
$$g_3^c=-{g_s\over\Lambda}\l({\sqrt{3}\over 2}s'+{2\over\sqrt{3}}s_1+\sqrt{3}
s_2-\sqrt{3}s_4-{1\over \sqrt{3}}s_5+{2\over \sqrt{3}}s_6\r),\eqno(3.52b)$$
$$g_3^{c*}=-{g_s\over\Lambda}\l({\sqrt{3}\over 2}s'+{2\over\sqrt{3}}s_1+\sqrt{
3}s_3+\sqrt{3}s_4+{2\over \sqrt{3}}s_5-{1\over \sqrt{3}}s_6\r),\eqno(3.52c)$$
$$g_5^c=-{g_s\over\Lambda}\l(-{3\over 2}s'+s_1+s_5+s_6\r),\eqno(3.52d)$$
where we have added a
term proportional to $s'$ due to the operator $O_1$ which
preserves the spin symmetry. As in the heavy meson case, we will assume that
the phases for the field operators of the heavy baryons have been so chosen
that all the coupling constants are real. Then, $g_3^c=g_3^{c*}$ gives
$$s_2-s_4+s_6=s_3+s_4+s_5.\eqno(3.53)$$
We can rewrite (3.52) in terms of the combinations
$$s_2'=s_2+s_3,\eqno(3.54a)$$
$$s_3'=s_5+s_6.\eqno(3.54b)$$
Finally, we obtain
$$g_1^c=-{g_s\over\Lambda}\l(s'+{1\over 3}s_1-s_2'+{4\over 3}s'_3\r),
\eqno(3.55a)$$
$$g_3^c=g_3^{c*}=-{g_s\over \Lambda}\l({\sqrt{3}\over 2}s'
+{2\over\sqrt{3}}s_1+
{\sqrt{3}\over 2}s_2'+{1\over 2\sqrt{3}}s_3'\r),\eqno(3.55b)$$
$$g_5^c=-{g_s\over\Lambda}\l(-{3\over 2}s'+s_1+s'_3\r).\eqno(3.55c)$$
   Eq.(3.55) shows that the spin symmetry relations (3.6a)-(3.6c) are
completely broken at order $O(1/m_Q)$ due to the presence of the parameters
$s_1,~s_2'$ and $s_3'$.

    We now turn to the $1/m_Q$ corrections to the radiative interactions
${\cal L}_{v,B}^{(2)}$, we shall treat the heavy and light quark
electromagnetic currents separately. It is known that the magnetic
couplings $a_1^{\prime}$-$a_6^{\prime}$, induced by heavy quark
electromagnetic currents, can be rigorously determined by the heavy quark
effective theory [15,17]. As in the meson case, coefficients $a_1^{\prime}$-
$
a_6^{\prime}$ arise entirely from the magnetic moment part of $J_{\mu}^{em}$
shown in Eq.(2.26). For antitriplet
baryons, we evaluate the magnetic form factor
$$G_{\bar{3}\mu}^Q=\bra {B_{\bar 3}(v,s')}{ie{\cal Q'}\over 2m_Q}k^{\nu}
\bar{h}_v\sigma_{\mu\nu}h_v\ket {B_{\bar 3}(v,s)}.\eqno(3.56)$$
Notice that we have taken $v=v'$ while maintaining a finite photon momentum
$k^{\nu}$. Applying Eq.(3.8) yields
$$G_{\bar{3}\mu}^Q={ie{\cal Q'}\over 2m_Q}k^{\nu}\bar{u}(v,s'){1+v\!\!\!/\
\over 2}\sigma_{\mu\nu}{1+v\!\!\!/\ \over 2}u(v,s)\bra {0}\phi_v \phi_v^{
\dagger}\ket {0},\eqno(3.57)$$
with [25]
$$\bra {0}\phi_{v}\phi_{v'}^{\dagger}\ket {0}=\zeta(v\cdot v'),\eqno(3.58)$$
and $\zeta(1)=1$.
After simplifying the gamma matrices and comparing the result with that given
by ${\cal L}_{v,B}^{(2)}$, we obtain
$$a_6^{\prime}=-{1\over 4}\left({e\over 2m_Q}\right).\eqno(3.59)$$
For magnetic transitions between sextet and antitriplet baryons, we evaluate
the following matrix element:
$$G_{6\bar{3}\mu}^Q=\bra {B_{\bar 3}(v,s')}{ie{\cal Q'}\over 2m_Q}k^{\nu}
\bar{h}_v\sigma_{\mu\nu}h_v\ket {B_6(v,s,\kappa)}.\eqno(3.60)$$
Application of Eqs.(3.8) and (3.29) yields
$$G_{6\bar{3}\mu}^Q={ie{\cal Q'}\over 2m_Q}k^{\nu}\bar{u}(v,s')
{1+v\!\!\!/\ \over 2}\sigma_{\mu\nu}{1+v\!\!\!/\ \over 2}B^{\alpha}(v,s,
\kappa)\bra {0}\phi_v \phi_{v,\alpha}^{\dagger}\ket {0}.\eqno(3.61)$$
Since the diquark fields $\phi$ and $\phi_{\mu}$ are scalar and axial vector
fields respectively, the matrix elements $ \bra {0}\phi_v \phi_{v,\alpha}^{
\dagger}\ket {0}$ must vanish due to  conservation of parity. This renders
$$a_2^{\prime}=0, \ \ \ a_4^{\prime}=0.\eqno(3.62)$$
Finally, we evaluate the following matrix elements to determine the
couplings $a_1^{\prime}$, $a_3^{\prime}$ and $a_5^{\prime}$:
$$G_{66\mu}^Q=\bra {B_6(v,s',\kappa^{\prime})}{ie{\cal Q'}\over 2m_Q}k^{\nu}
\bar{h}_v\sigma_{\mu\nu}h_v\ket {B_6(v,s,\kappa)}.\eqno(3.63)$$
Applying Eq.(3.29), we obtain
$$G_{66\mu}^Q={ie{\cal Q'}\over 2m_Q}k^{\nu}\bar{B}^{\alpha}(v,s',\kappa^{
\prime})
{1+v\!\!\!/\ \over 2}\sigma_{\mu\nu}{1+v\!\!\!/\ \over 2}B^{\beta}(v,s,\kappa)
\bra {0}\phi_{v,\alpha} \phi_{v,\beta}^{\dagger}\ket {0},\eqno(3.64)$$
with [25]
$$\bra {0}\phi_{v,\alpha} \phi_{v',\beta}^{\dagger}\ket {0}=-g_{\alpha\beta}
\xi_1(v\cdot v')+v^{\prime}_{\alpha}v_{\beta}\xi_2(v\cdot v')+\cdots,
\eqno(3.65)$$
where terms proportional
to $v_\alpha$ or/and $v'_\beta$ are not shown, as they
do not contribute to (3.64). The normalization of $\xi_1$ is given by
$$\xi_1(v\cdot v'=1)=1.\eqno(3.66)$$
In the $v=v'$ limit, the function $\xi_2$ does not contribute to $G_{66
\mu}^Q$ because
$$v^{\mu}B_{\mu}=0.\eqno(3.67)$$
One can explicitly work out $G_{66\mu}^Q$ by substituting Eq.(3.30)
into Eq.(3.64). Comparing results obtained in this manner with those
given by the relevant couplings in ${\cal L}_{v,B}^{(2)}$, we arrive at
$$a_1^{\prime}={1\over 6}\left({e\over 2m_Q}\right), \ \ a_3^{\prime}=
{-1\over \sqrt 3}\left({e \over 2m_Q}\right), \ \  a_5^{\prime}={1\over 2}
\left({e\over 2m_Q}\right).\eqno(3.68)$$
The results (3.59), (3.62) and (3.68) agree with the quark model calculations
[17].

   Next we tackle the light-quark electromagnetic currents, which give rise
to the magnetic couplings $a_1$-$a_6$. In the heavy quark mass expansion, we
again write
$$a_i=a_i^0+a_i^c{\Lambda\over 2m_Q},\eqno(3.69)$$
where $i=1,2,\cdots,6$. The relations among $a_i^0$'s were derived in Ref.[17]
 by evaluating the matrix elements  for $B_{\bar 3}\rightarrow B_{\bar 3}+
\gamma$, $B_6(B_6^*)\rightarrow B_{\bar 3}+\gamma$ and $B_6( B_6^*)\rightarrow
 B_6( B_6^*) +\gamma$. We find
$$a_3^0=-{\sqrt 3\over 2}a_1^0, \ \ \ a_5^0=-{3\over 2}a_1^0,
\ \ \ a_4^0={\sqrt 3}a_2^0, \ \ \ a_6^0=0.\eqno(3.70)$$
We shall follow the previous procedure to obtain the sub-leading
contributions. For $B_{\bar 3}\rightarrow B_{\bar 3}+\gamma$, we have
$$G_{{\bar 3}\mu}=\bra {B_{\bar 3}(v,s')}iT\int d^4x [O_1(x)+O_2(x)]j_{\mu}^
{em}(0)\ket {B_{\bar 3}(v,s)}_m,\eqno(3.71)$$
where the subscript $m$ indicates magnetic contributions only. As in the
previous section, we shall set the charge matrix ${\cal Q}={\bf 1}$ and
suppress all the flavor quantum numbers in the subsequent discussions.

Since the operator $O_1$ does not alter the Lorentz structure of light
constituents' matrix element, the magnetic form factors in Eq.(3.71)
receive no contributions from $O_1$ [17]. The contribution from $O_2$ is
given by
$$\eqalign{G_{{\bar 3}\mu}^2 &=-{g_s\over 4m_Q}
\bra {B_{\bar 3}(v,s')}iT\int d^4x \bar{h}_v\sigma_{\alpha\beta}G^{\alpha
\beta}h_v j_{\mu}^{em}(0)\ket {B_{\bar 3}(v,s)}_m\cr
&=-{g_s\over 4m_Q}\bar{u}(v,s')\sigma^{\alpha\beta}u(v,s)(M_{\alpha\beta\mu}^{
\bar 3})_m,\cr}\eqno(3.72)$$
where
$$M_{\alpha\beta\mu}^{\bar 3}=\bra {0}iT\int d^4x\,\phi_vG_{\alpha\beta}j_{
\mu}^{em}\phi_v^{\dagger}\ket {0}.\eqno(3.73)$$
Since $M_{\alpha\beta\mu}^{\bar 3}$ must be antisymmetric with respect to
the indices $\alpha$, $\beta$ and obeys constraints from both parity and
electromagnetic current conservation, one concludes
$$M_{\alpha\beta\mu}^{\bar 3}=-i\delta(g_{\mu\alpha}k_{\beta}
-g_{\mu\beta}k_{\alpha}),\eqno(3.74)$$
where $\delta$ is a constant independent of the heavy quark mass and $k$ is
the photon's momentum. In parametrizing $M_{\alpha\beta\mu}^{\bar 3}$, we
restrict ourselves to the structures linear in $k$ since
we are only interested in magnetic interactions.
The same simplification will be assumed in the subsequent discussions.
Substituting Eq.(3.74) into Eq.(3.72), we arrive at
$$G_{{\bar 3}\mu}^2={i\delta g_s\over 2m_Q}\bar{u}(v,s')
\sigma_{\mu\nu}k^{\nu}u(v,s).\eqno(3.75)$$
Eq.(3.75) corresponds to a change in the transition amplitude for $B_{\bar{3}}
\ri B_{\bar 3}+\gamma$:
$$\eqalign{\Delta\Gamma(B_{\bar 3}\ri B_{\bar 3}+\gamma) &=\bra{B_{\bar 3}\,
\gamma(k,\varepsilon)}iT\int d^4x O_2(x)
[-j_\mu^{em}(0)A^\mu(0)]\ket{B_{\bar 3}
}   \cr
&={g_s\delta\over 4m_Q}\,\bar{u}\sigma_{\mu\nu}F^{\mu\nu}u,  \cr}
\eqno(3.76)$$
where
$$F_{\mu\nu}\equiv i(k_\mu\varepsilon_\nu-k_\nu\varepsilon_\mu).\eqno(3.77)$$
Comparing Eq.(3.76) with ${\cal L}_{v,B}^{(2)}$, we conclude that
$$a_6^c={g_s\delta\over 2\Lambda}.\eqno(3.78)$$
Since $a_6^0=0$, the amplitude for the magnetic transition $B_{\bar 3}
\rightarrow B_{\bar 3}+\gamma$ is suppressed by $\Lambda/ 2m_Q$.
For $B_6(B_6^*)\rightarrow B_{\bar 3}+\gamma$, the
operator $O_2$ gives a contribution to the electromagnetic form factor
$$\eqalign{G_{6{\bar 3}\mu}^2 &=-{g_s\over 4m_Q}
\bra {B_{\bar 3}(v,s')}iT\int d^4x \bar{h}_v\sigma_{\alpha\beta}G^{\alpha
\beta}h_v j_{\mu}^{em}(0)\ket {B_6(v,s,\kappa)}\cr
&=-{g_s\over 4m_Q}\bar{u}(v,s')\sigma^{\alpha\beta}\bar{B}^{\nu}(v,s,\kappa)
(M_{\alpha\beta\mu\nu})_m,\cr}\eqno(3.79)$$
where
$$M_{\alpha\beta\mu\nu}=\bra {0}iT\int d^4x\,\phi_vG_{\alpha\beta}j_{\mu}^{
em}\phi_{v,\nu}^{\dagger}\ket {0}.\eqno(3.80)$$
The most general structure of $M_{\alpha\beta\mu\nu}$ relevant to the magnetic
transition is given by
$$M_{\alpha\beta\mu\nu}=
(g_{\alpha\nu}\epsilon_{\mu\beta\lambda\kappa}-g_{\beta
\nu}\epsilon_{\mu\alpha\lambda\kappa})k^\lambda v^\kappa t,\eqno(3.81)$$
where $t$ is a constant independent of the heavy quark mass.
Eqs.(3.79) and (3.81) give
$$\eqalign{\Delta\Gamma^2(6^*\ri \bar{3}) &= \bra{B_{\bar 3}\,\gamma(k,
\varepsilon)}iT\int d^4x O_2(x)[-j_\mu^{em}(0)A^\mu(0)]\ket{B_6^*}  \cr
&= {g_st\over 4m_Q}\epsilon_{\mu\nu\lambda\kappa}\bar{u}\gamma^\nu F^{\lambda
\kappa}u^\mu,  \cr}\eqno(3.82a)$$
$$\Delta\Gamma^2(6\ri \bar{3})=-{g_st\over 2m_Q}\,{1\over\sqrt{3}}\,\bar{u}_{
\bar 3}\sigma_{\mu\nu}F^{\mu\nu}u_6.\eqno(3.82b)$$
Comparing Eq.(3.82) with ${\cal L}_{v,B}^{(2)}$, we find
$$a_2^c=\,{g_s\over \Lambda}\l(t'-{t\over\sqrt{3}}\r),\eqno(3.83a)$$
$$a_4^c=\,{g_s\over \Lambda}\l(\sqrt{3}t'+{t\over 2}\r),\eqno(3.83b)$$
where
the spin-symmetry-preserving contributions proportional to $t'$ come from
the operator $O_1$.

  Finally we consider the couplings $a_1^c$, $a_3^c$ and $a_5^c$, which are
relevant to magnetic transitions among sextet baryons. The relevant matrix
element is
$$G_{66\mu}=\bra {B_6(v',s',\kappa^{\prime})}
iT\int d^4x [O_1(x)+O_2(x)]j_{\mu}
^{em}(0)\ket {B_6(v,s,\kappa)}_m.\eqno(3.84)$$
Particularly, we shall focus on the contribution from the operator $O_2$,
which is given by
$$\eqalign{G_{66\mu}^2 &=
-{g_s\over 4m_Q}\bra {B_6(v,s',\kappa^{\prime})}iT\int
d^4x \bar{h}_v\sigma_{\alpha\beta}G^{\alpha\beta}h_vj_{\mu}^{em}(0)\ket {B_6
(v,s,\kappa)}_m  \cr
&=-{g_s\over 4m_Q}\bar{B}^{\nu}(v,s',\kappa^{\prime})\sigma^{\alpha\beta}B^{
\lambda}(v,s,\kappa)(M_{\alpha\beta\nu\lambda\mu})_m,\cr}\eqno(3.85)$$
where
$$M_{\alpha\beta\nu\lambda\mu}=\bra {0}\phi_{v,\nu}G_{\alpha\beta}j_{\mu}^{
em}\phi_{v,\lambda}^{\dagger}\ket {0}.\eqno(3.86)$$
One can parametrize $M_{\alpha\beta\nu\lambda\mu}$ as follows:
$$\eqalign{ M_{\alpha\beta\nu\lambda\mu} &=iw_1g_{\nu\lambda}(g_{\mu\alpha}
k_{\beta}-g_{\mu\beta}k_{\alpha})\cr
&+iw_2[g_{\alpha\nu}(g_{\beta\mu}k_\lambda
-g_{\lambda\mu}k_\beta)-g_{\beta\nu}(g_{\alpha\mu}k_\lambda-g_{\lambda\mu}
k_\alpha)]  \cr
&+iw_3[g_{\alpha\lambda}(g_{\beta\mu}k_\nu
-g_{\mu\nu}k_\beta)-g_{\beta\lambda}(g_{\alpha\mu}k_\nu-g_{\mu\nu}k_\alpha)],
  \cr}\eqno(3.87)$$
where $w_1,~w_2$ and $w_3$ are constants independent of the heavy quark
mass. Eqs.(3.85) and (3.87) give a contribution to the photon transition
amplitude
$$\eqalign{ &\Delta\Gamma^2[B_6(v,s,\kappa) \ri B_6(v',s',\kappa')+\gamma(k,
\varepsilon)]  \cr
&=-\bra{B_6(v',s',\kappa')\gamma(k,\varepsilon)}iT\int d^4x
O_2(x)j_\mu^{em}(0)A^\mu(0)\ket{B_6(v,s,\kappa)}  \cr
&=-{g_s\over 4m_Q}\bar{B}_\nu(v',s',\kappa')(w_1\sigma^{\alpha\beta}F_{
\alpha\beta}g^{\nu\lambda}+2w_2\sigma^{\nu\beta}F^{\ \lambda}_\beta+2w_3
\sigma^{\lambda\beta}F^{\ \nu}_\beta)B_\lambda(v,s,\kappa).  \cr}\eqno(3.88)$$
Let us denote
$$\Delta\Gamma^2=-{g_s\over 4m_Q}(\delta_1+\delta_2+\delta_3),\eqno(3.89)$$
where
$$\delta_1=w_1\bar{B}^\nu(v',s',\kappa')\sigma^{\alpha\beta}F_{\alpha\beta}
B_\nu(v,s,\kappa),\eqno(3.90a)$$
$$\delta_2=2w_2\bar{B}_\nu(v',s',\kappa')\sigma^{\nu\beta}F_{\beta\lambda}
B^\lambda(v,s,\kappa),\eqno(3.90b)$$
$$\delta_3=2w_3\bar{B}^\nu(v',s',\kappa')\sigma^{\lambda\beta}F_{\beta\nu}
B_\lambda(v,s,\kappa).\eqno(3.90c)$$
Making uses of the identities for the Dirac matrices stated earlier and
$$\{\sigma_{\mu\nu},~\gamma_\lambda\}=2\epsilon_{\mu\nu\lambda\kappa}\gamma
^\kappa\gamma_5,\eqno(3.91a)$$
$$[\sigma_{\mu\nu},~\gamma_\lambda]=
-2i(g_{\mu\lambda}\gamma_\nu-g_{\nu\lambda}
\gamma_\mu),\eqno(3.91b)$$
$$\gamma^\lambda\sigma_{\mu\nu}\gamma_\lambda=0,\eqno(3.91c)$$
$$\bar{u}^\lambda(v)\sigma_{\mu\nu}F^{\mu\nu}u_\lambda(v)=2i\bar{u}^\mu(v)
F_{\mu\nu}u^\nu(v),\eqno(3.91d)$$
we obtain
$$\delta_1(6^*\ri 6^*)=w_1\bar{u}^\nu\sigma^{\alpha\beta}F_{\alpha\beta}u_\nu,
\eqno(3.92a)$$
$$\delta_1(6^*\ri 6)=-{2\over\sqrt{3}}w_1\epsilon_{\mu\nu\lambda\kappa}\bar{u}
\gamma^\nu F^{\lambda\kappa}u^\mu,\eqno(3.92b)$$
$$\delta_1(6\ri 6^*)=-{2\over\sqrt{3}}w_1\epsilon^{\mu\nu\lambda\kappa}\bar{u}
_\mu\gamma_\nu F_{\lambda\kappa}u,\eqno(3.92c)$$
$$\delta_1(6\ri 6)={1\over 3}w_1\bar{u}\sigma^{\alpha\beta}F_{\alpha\beta}u,
\eqno(3.92d)$$

$$\delta_2(6^*\ri 6^*)=
-w_2\bar{u}^\nu\sigma^{\alpha\beta}F_{\alpha\beta}u_\nu,
\eqno(3.93a)$$
$$\delta_2(6^*\ri 6)={2\over\sqrt{3}}w_2\epsilon_{\mu\nu\lambda\kappa}\bar{u}
\gamma^\nu F^{\lambda\kappa}u^\mu,\eqno(3.93b)$$
$$\delta_2(6\ri 6^*)=-{1\over\sqrt{3}}w_2\epsilon^{\mu\nu\lambda\kappa}\bar{u}
_\mu\gamma_\nu F_{\lambda\kappa}u,\eqno(3.93c)$$
$$\delta_2(6\ri 6)=-{4\over 3}w_2\bar{u}\sigma^{\alpha\beta}F_{\alpha\beta}u,
\eqno(3.93d)$$

$$\delta_3(6^*\ri 6^*)=-w_3\bar{u}^\nu\sigma^{\alpha\beta}
F_{\alpha\beta}u_\nu,
\eqno(3.94a)$$
$$\delta_3(6^*\ri 6)=-{1\over\sqrt{3}}w_3\epsilon_{\mu\nu\lambda\kappa}\bar{u}
\gamma^\nu F^{\lambda\kappa}u^\mu,\eqno(3.94b)$$
$$\delta_3(6\ri 6^*)={2\over\sqrt{3}}w_3\epsilon^{\mu\nu\lambda\kappa}\bar{u}
_\mu\gamma_\nu F_{\lambda\kappa}u,\eqno(3.94c)$$
$$\delta_3(6\ri 6)=-{4\over 3}w_3\bar{u}\sigma^{\alpha\beta}F_{\alpha\beta}u,
\eqno(3.94d)$$
Collecting all the terms yields
$$\Delta\Gamma^2(6^*\ri 6^*)=-{g_s\over 4m_Q}(w_1-w_2-w_3)\bar{u}^\nu\sigma^{
\alpha\beta}F_{\alpha\beta}u_\nu,\eqno(3.95a)$$
$$\Delta\Gamma^2(6^*\ri 6)=
-{g_s\over 4m_Q}\l(-{2\over\sqrt{3}}w_1+{2\over\sqrt
{3}}w_2-{1\over\sqrt{3}}w_3\r)\epsilon_{\mu\nu\lambda\kappa}\bar{u}
\gamma^\nu F^{\lambda\kappa}u^\mu,\eqno(3.95b)$$
$$\Delta\Gamma^2(6\ri 6^*)=
-{g_s\over 4m_Q}\l(-{2\over\sqrt{3}}w_1-{1\over\sqrt
{3}}w_2+{2\over\sqrt{3}}w_3\r)\epsilon_{\mu\nu\lambda\kappa}\bar{u}^\mu
\gamma^\nu F^{\lambda\kappa}u,\eqno(3.95c)$$
$$\Delta\Gamma^2(6\ri 6)=-{g_s\over 4m_Q}\l({1\over 3}w_1-{4\over 3}w_2-{4
\over 3}w_3\r)\bar{u}\sigma^{\alpha\beta}F_{\alpha\beta}u.\eqno(3.95d)$$
When Eq.(3.95) is compared with $\L^{(2)}_{v,B}$, we have
$$a_1^c=-{g_s\over 2\Lambda}\l[w+{1\over 3}(w_1-4w_2-4w_3)\r],\eqno(3.96a)$$
$$a_3^c=-{g_s\over 2\Lambda}\l[-{\sqrt{3}\over 2}w+{1\over \sqrt{3}}(-2w_1-w_2
+2w_3)\r],\eqno(3.96b)$$
$$a_3^{c*}=-{g_s\over 2\Lambda}\l[-{\sqrt{3}\over 2}w+{1\over \sqrt{3}}(-2w_1+
2w_2-w_3)\r],\eqno(3.96c)$$
$$a_5^c=-{g_s\over 2\Lambda}\l[-{3\over 2}w+(w_1-w_2-w_3)\r],\eqno(3.96d)$$
where we have included the contributions proportional to $w$ from the spin
symmetry preserving operator $O_1$. Again, we will assume that the phases of
the heavy baryon fields have been so chosen that the coupling constants are
real. Then $a_3^c=a_3^{c*}$ gives
$$w_2=w_3,\eqno(3.97)$$
and
$$a_1^c=-{g_s\over 2\Lambda}\l(w+{1\over 3}w_1-{8\over 3}w_2\r),\eqno(3.98a)$$
$$a_3^c=a_3^{c*}=
-{g_s\over 2\Lambda}\l(-{\sqrt{3}\over 2}w-{2\over\sqrt{3}}w_1
+{1\over\sqrt{3}}w_2\r),\eqno(3.98b)$$
$$a_5^c=-{g_s\over 2\Lambda}\l(-{3\over 2}w+w_1-2w_2\r),\eqno(3.98c)$$
It is clear from Eqs.(3.78), (3.83) and (3.98) that to order $O(1/m_Q)$
all the spin symmetry relations among the coupling constants $a_i$ for
radiative transitions are broken.

   Eqs.(3.16), (3.36) and (3.55) are the main results in this section for the
strong coupling constants
$g_1,\cdots,g_6$, and (3.78), (3.83), (3.98) for the
electromagnetic transition couplings $a_1,\cdots,a_6$.

\endpage

\noindent{\bf 4. Discussion  }
\vskip 0.3 cm
   In this work we have carried out a systematic theoretical study of the
order $1/m_Q$ effects to the heavy meson's and heavy baryon's chiral
Lagrangian for strong and electromagnetic interactions. There are two distinct
corrections at this order. The first is a kinematic correction required by
reparametrization invariance. In practice, this effect for simple processes
such as decays can be largely
taken into account by using the full momentum $P$
of a heavy particle and the corresponding polarization vector or Dirac spinor
instead of parametrizing it by $P=M_Hv+k$ and dropping
the residual momentum $k$. The second effect is a dynamic correction induced
by the order $1/m_Q$ terms in the QCD Lagrangian which break the flavor-spin
symmetry of the heavy quarks. As in our earlier publications [11,17],
we focus our attention on the interactions involving only the first order
in the momentum of a Goldstone boson or a photon. To this order, not
surprisingly, the heavy quark symmetry breaking interactions of QCD do not
produce any new types of interactions for the heavy hadrons with the Goldstone
bosons or photons. Instead, their effects make order $1/m_Q$ corrections to
the coupling constants in the heavy hadron's chiral Lagrangian.

   To order $1/m_Q$, QCD contains one operator $O_1$ [see Eq.(1.15c)] which
breaks only the heavy flavor symmetry, and a second operator $O_2$ [see
Eq.(1.15d)] which breaks the flavor-spin symmetry of heavy quarks. For
a given heavy flavor, the effects due to $O_1$ can be absorbed by the coupling
constants which satisfy the heavy quark spin symmetry. Effectively, the
operator $O_1$ does not introduce any new unknowns. On the other hand, the
operator $O_2$ introduces new unknowns of order $1/m_Q$ which break the
spin symmetry relations among the coupling constants. In the heavy meson
sector, there is one new unknown each in the strong interactions and
electromagnetic interactions, respectively. In the heavy baryon sector, there
are five new unknowns of order $1/m_Q$ to describe the six
strong interaction coupling constants $g_1,\cdots,g_6$. There are four new
unknowns to describe the six radiative transition coupling
constants $a_1,\cdots,a_6$. In particular, the reactions $B_{\bar 3}\ri
B_{\bar 3}+\pi$ and $B_{\bar 3}\ri B_{\bar 3}+\gamma$ which are forbidden in
the infinitely heavy quark limit have coupling strength of order $1/m_Q$.
In reducing the number of unknowns, we have appealed to the charge conjugation
symmetry for certain processes and the reality of the coupling constants
associated with them.
This can always be accomplished by a proper choice of the
phases for the field operators of the heavy hadrons. For example, in our
quark model calculations in Refs.[11,17], all the coupling constants are
indeed real.

   These new unknowns depend on the QCD's long distance dynamics of light
quarks and gluons. In principle, they are calculable numerically in lattice
QCD.
At a more pheomenological level, the quark model has no simple predictions
for them either, unlike the coupling constants in the infinitely heavy quark
limit.
Nevertheless, it is important to consider the sizes of these corrections
as they affect the strong and electromagnetic interaction physics of the
heavy hadrons, especially the charmed mesons and baryons. For this purpose, it
is perhaps useful to calculate those corrections in the quark model and the
MIT bag model with some specific pheomenological wave functions for the heavy
hadrons.

   We will illustrate the last point by a problem of practical interest.
For the heavy meson chiral Lagrangian $\L^{(1)}_{v,PP^*}$, the HQS relation
$f=2g$ is modified by $1/m_Q$ corrections.
The splitting of $f$ and $2g$, $\delta\equiv 2g-f$, will contribute to
$SU(3)$-violating corrections to the heavy-meson hyperfine splitting. Such
corrections are characterized by the parameter [26]
$$\Delta_P\equiv (M_{P_s^*}-M_{P_s})-(M_{P_d^*}-M_{P_d}).$$
In the charmed meson case, experimental data give [27,28]:
$$\Delta_D=0.9\pm1.9 \ {\rm MeV}.\eqno(4.1)$$
On the theoretical side, a one-loop calculation based on the heavy-meson
chiral Lagrangian given by Eq.(2.1) has recently been performed [23].
In this work $\Delta_D$ is obtained by evaluating the self-energy diagrams of
$D$ and $D^*$, where each diagram contains one insertion of the residual mass
term $\Delta M^2P(v)P^{\dagger}(v)$ appearing in ${\cal L}_{v,PP^*}^{(1)}$.
By retaining $m_s\ln m_s$ and $m_s^{3/2}$ corrections, it was found that, to
the order of $1/m_c$ [23],
$$\Delta_D=+95 \ {\rm MeV}.\eqno(4.2)$$
In comparison with the experimental value given by Eq.(4.1), this result is
larger by almost two orders of magnitude! In addition to obtaining Eq.(4.2),
the authors of Ref.[23] also estimated other contributions to $\Delta_D$,
which are of the same order or one order higher.
As they pointed out, one had to include an additional contribution which is
quadratic in Goldstone boson masses.  Such corrections arise from the
chiral-loops mentioned above plus the counterterms listed below (see also
Ref.[1]):
$$\eqalign{&{\cal O}^P=\alpha_1M_PP(\xi {\cal M}^{\dagger}\xi+ \xi^{\dagger}
{\cal M}\xi^{\dagger} )P^{\dagger},\cr
&{\cal O}^{P^*}=\alpha_2M_{P^*}P^*_{\mu}(\xi {\cal M}^{\dagger}\xi+ \xi^{
\dagger} {\cal M}\xi^{\dagger} )P^{*\mu\dagger},\cr}\eqno(4.3)$$
where
$${\cal M}=\pmatrix {m_u&0&0\cr 0&m_d& 0\cr
0&0&m_s\cr}.\eqno(4.4)$$
The counterterms ${\cal O}^P$ and ${\cal O}^{P^*}$ contribute to $\Delta_D$
because there will be a deviation from the spin-symmetry relation
$\alpha_1=-\alpha_2$ at the order of $1/m_c$. By a naive dimensional argument,
Randall and Sather estimated this contribution to be about $20 \ {\rm MeV}$
in magnitude. Furthermore, they noted that the $O(1/m_c^2)$ contributions to
$\Delta_D$ could also be as large as $10 \ {\rm MeV}$ in magnitude.  These
large individual corrections raise interesting questions on the reliability
of chiral perturbation theory and/or $1/m_Q$ expansion. Given all
these large contributions, the authors suggested that there may be accidental
cancellations among various terms so that the resultant $\Delta_D$ is small.
Although this might indeed be the case, we would like to point out that the
calculation done in Ref.[23] is not complete at order $1/m_c$.
Specifically it missed those effects induced by the splitting of $f$ and $2g$
in the self-energy diagram depicted in
Fig.1. At the order of $1/m_c$, $\delta$ is finite, and
$D$ and $D^*$ will acquire different mass-shifts which
contributes to the hyperfine splittings of heavy mesons. Since the difference
between $D$ and $D^*$ mass-shifts is $SU(3)$ flavor dependent, it therefore
contributes to the parameter $\Delta_ D$. Denoting this extra contribution as
$\Delta_D^{\prime}$, we find
$$\Delta_D^{\prime}=\left({1\over 32}{m_{\pi}^3\over \pi f_{\pi}^2}
-{1\over 48}{m_{_K}^3\over\pi f_{_K}^2}-{1\over 96}{m_{\eta}^3
\over \pi f_{\eta}^2}\right)x -{3\over 32}{m_{_K}^2\over \pi^2 f_{_K}^2}
\ln\l({m_{_K}^2\over \Lambda_{\chi}^2}\r)(M_{D_s}-M_{D_d})x,\eqno(4.5)$$
where $x=f\delta$, and we have suppressed contributions proportional to $(M_{
D_s}-M_{D_d})^2$ or higher since they are found to be negligible.
The first term in Eq.(4.5) can be easily obtained from Eq.(4.3) of our
forthcoming paper [1] (see the footnote there for details). The second term
emerges as one takes into account the splitting of the strange and non-strange
 heavy-meson masses. Now for numerical analyses, we shall take $\Lambda_{\chi}
\approx 1$ GeV [29], and the fitted value of $(M_{D_s}-M_{D_d})= 99.5\pm 0.6
\ {\rm MeV}$ [27].
In view of large positive value in Eq.(4.2), one would favor
a negative $\Delta_D^{\prime}$ to counteract it. This requires $x$ to be
positive or, in other words, $4g^2>f^2$. At this point,  we do not plan to
perform any model-calculation of $x$. Nevertheless, a crude estimation of $x$
can be obtained by dimensional arguments. If we assume that the
heavy quark expansion at the hadronic level is governed by
the parameter $\Lambda_{\chi}/2m_Q$ [22], we would roughly expect that
$$\left|{\delta\over f}\right|\approx {\cal O}\l({\Lambda_{\chi}
\over 2m_c}\r).
\eqno(4.6)$$
Taking $m_c=1.8 \ {\rm GeV}$ and $f^2=2$ [30], we
obtain
$${|x|\over f^2}=\left|{\delta\over f}\right|\approx  {\cal O}(0.3).
\eqno(4.7)$$
If one simply assumes $x=0.3 f^2=0.6$, then $\Delta_D^{\prime}=-62 \ {\rm
MeV}$. If $x$ is indeed positive, this would provide a substantial
cancellation to the result of Eq.(4.2).  The cancellation would be further
enhanced if the contribution from Eq.(4.3) is also negative. Unfortunately,
there still exists no data to support this claim. In fact, there is also no
experimental evidence for a positive $x$. Certainly, a negative $x$ would make
 the situation even more troublesome. At any rate, we want to emphasize that
one should include the effect of Fig. 1 when computing the parameter
$\Delta_D$. Whether or not this effect is adequate to resolve the puzzle
posed by Eq.(4.2) is not yet clear until one has more experimental data,
and a better theoretical understanding [31].

  From the above example, we have seen that the splitting of $f$ and $2g$ at
 the order of $1/m_Q$ could give important effects. Similar situations may
also occur in other parts of the heavy-hadron chiral Lagrangian. Since there
are insufficient data to fix the parameters of the theory, it would be
helpful to combine the results of this paper with certain model-estimation of
parameters, so that quantitative predictions of the $1/m_Q$ correction can be
made. We shall leave such model studies to future investigation.

\bigskip
\bigskip

\centerline{\bf Acknowledgments}
\bigskip

   We like to thank Dr. Peter Cho for a discussion concerning his work [18].
   One of us (H.Y.C.) wishes to thank Prof. C. N. Yang and the
Institute for Theoretical Physics at Stony Brook for their hospitality
during his stay there for sabbatical leave.
T.M.Y.'s work is supported in part by the
National Science Foundation.  This research is supported in part by the
National Science Council of ROC under Contract Nos.  NSC82-0208-M001-001Y,
NSC82-0208-M001-016, NSC82-0208-M001-060 and NSC82-0208-M008-012.

\endpage

\centerline{\bf REFERENCES}
\medskip

\item {1.}
 H.-Y.
Cheng, C.-Y. Cheung, G.-L. Lin, Y.-C. Lin, T.-M. Yan and H.-L. Yu,
CLNS 93/1189, IP-ASTP-01-93, ITP-SB-93-03 (1993).

\item {2.}
  N. Isgur and M. B. Wise, {\sl Phys. Lett.} {\bf B232}, 113 (1989); {\sl
Phys. Lett.} {\bf B237}, 527 (1990).

\item {3.}
  M. B. Voloshin and M. A. Shifman,
{\sl Yad. Fiz.} {\bf 45}, 463 (1987) [{\sl Sov. J. Nucl. Phys.} {\bf 45}, 292
(1987)].

\item {4.}
  E. Eichten and B. Hill,
{\sl Phys. Lett.} {\bf B234}, 511 (1990).

\item {5.}
  H. Georgi, {\sl Phys. Lett.} {\bf B240}, 447 (1990).

\item {6.}
  B. Grinstein, {\sl Nucl. Phys.}
{\bf B339}, 253 (1990).

\item {7.}
  E. Eichten and B. Hill, {\sl Phys. Lett.}
{\bf B243}, 427 (1990).

\item {8.}
  A. F. Falk, B. Grinstein, and M. Luke,
{\sl Nucl. Phys.} {\bf B357,} 185 (1991).

\item {9.}
  M. Luke, {\sl Phys.   Lett.} {\bf B252,} 447 (1990).

\item {10.}
  See Refs.[2,3] and others such as:
N. Isgur and M. B. Wise, {\sl Nucl. Phys.} {\bf B348}, 276 (1991); H. Georgi,
{\sl Nucl. Phys.} {\bf B348}, 293 (1991).

\item {11.}
  T.-M. Yan, H.-Y. Cheng, C.-Y. Cheung, G.-L. Lin, Y.-C. Lin,
and H.-L. Yu, {\sl Phys. Rev}. {\bf D46}, 1148 (1992). See also T.-M. Yan,
{\sl Chin. J. Phys. (Taipei)} {\bf 30}, 509 (1992).

\item {12.}
  M. B. Wise,
{\sl Phys. Rev.} {\bf D45}, R2188 (1992).

\item {13.}
  G. Burdman and
J. Donoghue, {\sl Phys. Lett}. {\bf B280}, 287 (1992).

\item {14.}
  P. Cho,
{\sl Phys. Lett}. {\bf B285}, 145 (1992).

\item {15.}
  P. Cho and H. Georgi,
{\sl Phys. Lett.} {\bf B296}, 408 (1992);
{\sl ibid} {\bf B300}, (E)410 (1993).

\item {16.}
  J. F. Amundson,
C. G. Boyd, E. Jenkins, M. Luke, A. V. Manohar,
J. L. Rosner, M. J. Savage, and
M. B. Wise, {\sl Phy. Lett}. {\bf B296}, 415 (1992).

\item {17.}
  H.-Y.
Cheng, C.-Y. Cheung, G.-L. Lin, Y.-C. Lin, T.-M. Yan, and H.-L. Yu, {\sl Phys.
 Rev}. {\bf D47}, 1030 (1993).

\item {18.}
  A special case of these
corrections was discussed by P. Cho [{\sl Nucl. Phys.} {\bf B396},
183 (1993)].
However, the result stated there is different from ours. The
difference between the two results will be discussed in Section 3.
Dr. Cho has
informed us that he has now corrected and clarified his result in a
preprint, CALT-68-1844 (1993).

\item {19.}
  E. Jenkins and A. V. Manohar, {\sl Phys. Lett}. {\bf B255}, 558 (1991);
E. Jenkins and A. V. Manohar, in {\it Proceedings of the Workshop on Effective
Field Theories of the Standard Model,}
edited by Ulf-G. Mei\ss ner; V. Bernard, N. Kaiser, J. Kambor and
Ulf-G. Mei\ss ner, BUTP-92/15, CRN 92-24, TUM-T31-28/92.

\item {20.}
  M. Luke and A. V. Manohar, {\sl Phys. Lett.} {\bf B286}, 348 (1992).

\item {21.}
See the discussion in Section 3 on the $B_{\bar 3}B_{\bar 3}\A$ coupling and
its relation to Cho's work [18] and see also  M. Neubert, {\sl Phys. Lett.}
{\bf B306}, 357 (1993); A.F. Falk and M. Luke, {\sl ibid} {\bf B292}, 119
(1992).

\item {22.}
 L. Randall and E. Sather, MIT-CTP \# 2167 (1992).

\item {23.}
 L. Randall and E. Sather, {\sl Phys. Lett.} {\bf B303}, 345 (1993).

\item {24.}
  J. D. Bjorken, SLAC-PUB-5278, invited talk at
Recontre de Physique de la Vallee d'Acoste, La Thuile, Italy (1990)
unpublished.

\item {25.}
   See Georgi in Ref.[10].

\item {26.}
  J. Rosner and M. B. Wise, {\sl Phys.
 Rev}. {\bf D47}, 343 (1993).

\item {27.}
  Particle Data Group, {\sl Phys.
 Rev}. {\bf D45}, S1 (1992).

\item {28.}
CLEO Collaboration, F. Butler {\it et al}, {\sl Phys. Rev. Lett.}
{\bf 69}, 2041 (1992).

\item {29.}
  A. Manohar and H. Georgi, {\sl Nucl. Phys.} {\bf B234}, 189 (1984).

\item {30.}
We choose this value for $f^2$ because the result quoted in Eq.(4.1) were
obtained with an identical choice.

\item {31.}
For one possible theoretical approach to the issue, see B. Rosenstein and
H. L. Yu, IP-ASTP-05-93 (1993).

\vskip 0.6cm
\vskip 0.6cm

\centerline{\bf Figure Caption}
\vskip 0.6cm
\item{Fig.1}
An additional chiral-loop diagram which contributes to the hyperfine
splitting of charmed mesons.
The dotted line denotes a light meson which can be strange or non-strange.
The solid line represents a charmed meson which can be strange or
non-strange, spin zero or spin 1. In this case, the propagator
of the heavy meson contains no insertion of the mass-difference term,
$\Delta M^2P(v)P^{\dagger}(v)$.

\endpage

\end